\RequirePackage{fix-cm}
\documentclass[smallextended]{svjour3}       % onecolumn (second format)
\smartqed  % flush right qed marks, e.g. at end of proof
\usepackage{graphicx}
\usepackage{txfonts}
\usepackage{natbib}
\usepackage{mathptmx}      % use Times fonts if available on your TeX system
\usepackage{amssymb}	% Extra maths symbols
\newcommand{\vv}[1]{\vec{#1}}

\newcommand{\bcd}{\mathbf{\cdot}}
\newcommand{\bcr}{\mathbf{\times}}
\newcommand{\qq}[1]{\mathsf{#1}}
\newcommand{\rd}{\mathrm{d}}
\newcommand{\cH}{\mathcal{H}}
\newcommand{\cK}{\mathcal{K}}
\newcommand{\cJ}{\mathcal{J}}
\newcommand{\cM}{\mathcal{M}}

\journalname{CELE}

\begin{document}

\title{The Lissajous-Kustaanheimo-Stiefel transformation}

\author{
 Slawomir Breiter
 \and
 Krzysztof Langner}

\institute{S. Breiter \at Astronomical Observatory Institute, Faculty of Physics, Adam Mickiewicz
       University, Sloneczna 36, 61-286 Poznan, Poland \\
       \email{breiter@amu.edu.pl}
       \and
K. Langner \at Astronomical Observatory Institute, Faculty of Physics, Adam Mickiewicz
       University, Sloneczna 36, 61-286 Poznan, Poland \\
       \email{krzysztof.langner@amu.edu.pl}
       }

\date{Received: date / Accepted: date}

\maketitle

\begin{abstract}

The Kustaanheimo-Stiefel transformation of the Kepler problem with a time-dependent perturbation converts it
into a perturbed isotropic oscillator of 4-and-a-half degrees of freedom with additional constraint known as bilinear invariant.
Appropriate action-angle variables for the constrained oscillator are required to apply canonical perturbation techniques
in the perturbed problem. The Lissajous-Kustaanheimo-Stiefel (LKS) transformation is proposed, leading to the action-angle
set which is free from singularities of the LCF variables earlier proposed by Zhao. One of the actions is the bilinear invariant,
which allows the reduction back to the 3-and-a-half degrees of freedom. The transformation avoids any reference to the
notion of the orbital plane, which allowed to obtain the angles properly defined not only for most of the
circular or equatorial orbits, but also for the degenerate, rectilinear ellipses. The Lidov-Kozai problem is analyzed in terms of
the LKS variables, which allow a direct study of stability for all equilibria except the circular equatorial and the polar radial orbits.


\end{abstract}

\section{Introduction}

The Kustaanheimo-Stiefel (KS) transformation is probably the most renowned regularization technique for the three-dimensional Kepler problem.
In the planar case, the conversion of the Kepler problem into a harmonic oscillator has been known since \citet{Gours:89} and
\citet{LC:1906}, but its extension to the three-dimensional problem took many decades of futile efforts.
Finally, \citet{Kust:64} discovered that the way to the third dimension is not direct, but requires a detour
through a constrained problem with four degrees of freedom. The KS transformation gained popularity in the matrix-vector formulation of
\citet{KS:65}, but it is much easier to interpret and generalize in the language of quaternion algebra, very closely related with the original
spinor formulation of \citet{Kust:64}.

The most common use of the KS transformation is the numerical integration of perturbed elliptic motion, where many intricacies introduced by the additional degree of freedom can be ignored,
although -- as recently demonstrated by \citet{RUP:16} -- they can be quite useful in the assessment of a global integration error.
Analytical perturbation methods for  KS-transformed problems often follow the way indicated by \citet{KS:65} and developed by \citet{StS:71}:
 variation of arbitrary constants is applied to constant vector amplitudes of the KS coordinates and velocities. But those who want to benefit from the
wealth of canonical formalism, require a set of action-angle variables of the regularized Kepler problem.

The first step in this direction can be found in the monograph by \citet{StS:71}, where the symplectic polar coordinates are introduced for
each separate degree of freedom. However, this approach does not account for degeneracy of the problem and thus is unfit for the averaging-based perturbation techniques.
Moreover, no attempt was made to relate this set with the constraint known as the `bilinear invariant', effectively reducing the system to 3 degrees of freedom.
Both problems have been resolved by \citet{Zhao:15}, who proposed the `LCF' variables (presumably named after \citet{LC:1906} and \citet{Fej:2001}).
In his approach, the motion in the KS variables is considered in an osculating `Levi-Civita plane' \citep{DEF} as a two degrees of freedom problem.
The third degree of freedom is added by the pair of action-angle variables orienting the plane. The redundant fourth degree is hidden in the
definition of the Levi-Civita plane. The transformed Keplerian Hamiltonian depends on a single action variable, the other two actions being closely
related to the angular momentum and its projection on the polar axis. Interestingly, the result is  identical to the `isoenergetic variables'
found by \citet{LC:1913} without regularization.

The LCF variables respect the degeneracy and bring the oscillations back to three degrees of freedom. Yet they possess a significant weakness: they are founded
on the orientation of a plane determined by the angular momentum. Whenever the angular momentum vanishes (even temporarily), the angles become undetermined and
equations of motion are singular. It turns out that seeking the proximity to the Delaunay variables, \citet{Zhao:15} reintroduced the singularities of unregularized
Kepler problem. Of course, some singularities are inevitable when the problem having spherical topology is mapped onto a torus of action-angle variables.
But there is always some freedom in the choice of the singularities.  Recalling that the main purpose of regularization is to allow the study of highly elliptic and rectilinear
orbits, we find it worth an effort to construct the action-angle set that -- unlike the LCF variables -- is regular for this class of motions.

The main goal of the present work is to derive an alternative set of the action-angle variables which is not based upon the notion of an orbital plane (thus avoiding
singularities when the orbit degenerates into a straight segment), and to test it on some well known astronomical problem.
Section~\ref{KStr} introduces some preliminary concepts related to the KS coordinate transformation in the language of quaternions.
We use its generalized form with an arbitrary `defining vector' \citep{BL:17}, which helps to realize how the choice of the KS1 or KS3
convention  allows or inhibits the use of the Levi-Civita plane in the construction of the action-angle sets. We have also benefited from the opportunity
to polish and extend the geometrical interpretation given to the KS transformation by \citet{Saha:09}.
In Section~\ref{KScan} we complement the KS coordinates with their
conjugate momenta and provide the Hamiltonian function in the extended phase space as the departure point for further transformations.
Section~\ref{AAs} builds the new action-angle set -- the Lissajous-Kustaanheimo-Stiefel (LKS) variables. Two independent Lissajous transformations
are followed by a linear Mathieu transformation. In Section~\ref{interp} we show how to interpret the new variables
not only in terms of the Lissajous ellipses, but also  by the reference to the angular momentum and Laplace vectors
of the Kepler problem.
As an application, we discuss the classical Lidov-Kozai problem (Section~\ref{Koza}), showing that stability
of rectilinear orbits can be discussed directly in terms of the LKS variables, which has not been possible using the Delaunay or
the LCF framework. Conclusions and future prospects are presented in the closing Section~\ref{conc}.

\section{KS transformation in quaternion form}
\label{KStr}
\subsection{Quaternion algebra}

Adhering to the convention used by \citet{DEF}, we treat a quaternion $\qq{v} \in \mathbb{H}$
as union of a scalar $v_0$ and a vector $\vv{v}$,
\begin{equation}\label{qdef}
    \qq{v} = \left( v_0, \vv{v} \right) = \sum_{j=0}^3 v_j \qq{e}_j,
\end{equation}
where the standard basis quaternions
\begin{equation}\label{baza}
    \qq{e}_0 = (1,\vv{0}), \quad  \qq{e}_1 = (0,\vv{e}_1),  \quad \qq{e}_2 = (0,\vv{e}_2), \quad \qq{e}_3 = (0,\vv{e}_3),
\end{equation}
have been defined by referring to the standard vector basis $\vv{e}_j$. Downgrading a `pure quaternion' $\qq{u} = (0,\vv{u})  \in \mathbb{H}'$
to a vector $\vv{u} \in \mathbb{R}^3$ requires application of the projection operator $\natural$, whose action on any quaternion
is $\qq{v}^\natural = (v_0, \vv{v})^\natural = \vv{v}$.

As members of the Euclidean linear space $\mathbb{R}^4$, quaternions admit the sum and product-by-scalar rules
\begin{equation}\label{ssp}
    \qq{u} + \qq{v} = \sum_{j=0}^3 \left(u_j+v_j\right) \qq{e}_j, \qquad \alpha \qq{v} =  \sum_{j=0}^3 \alpha v_j \qq{e}_j,
\end{equation}
as well as the scalar product
\begin{equation}\label{sps}
    \qq{u} \bcd \qq{v} = \sum_{j=0}^3  u_j v_j  = u_0 v_0 + \vv{u} \bcd \vv{v},
\end{equation}
implying the norm
$ |\qq{v} | =   \sqrt{\qq{v} \bcd \qq{v} } = \sqrt{v_0^2 + \|\vv{v}\|^2}$, where $\|\vv{v}\| = \sqrt{\vv{v}\cdot \vv{v}},$
to distinguish the norms in $\mathbb{R}^3$ and $\mathbb{R}^4$.

What makes four-vectors $\qq{u}$ and $\qq{v}$ the members of the quaternion algebra $\mathbb{H}$ over $\mathbb{R}$,
is the noncommutative quaternion product definition
\begin{equation}\label{qp}
  \qq{u}\, \qq{v} = \left( u_0 v_0 - \vv{u} \cdot \vv{v} , u_0 \vv{v} + v_0 \vv{u} +  \vv{u} \bcr \vv{v}\right).
\end{equation}
Note that $\mathbb{H}'$ is only a linear subspace, but not a subalgebra of $\mathbb{H}$, because the quaternion product
of two pure quaternions may have a nonzero scalar part.

Two other useful operations to be defined are the quaternion conjugate
\begin{equation}\label{conq}
    \overline{\qq{v}} =  (v_0 , - \vv{v}),
\end{equation}
allowing to write $|\qq{v}|^2 = \qq{v} \overline{\qq{v}}$, and the quaternion cross product
\begin{equation} \label{qop}
\qq{u}  \wedge  \qq{v} = \frac{ \qq{v} \bar{\qq{u}} - \qq{u} \bar{\qq{v}} }{2}
= \left( 0, \, u_0 \vv{v} - v_0 \vv{u} + \vv{u} \bcr \vv{v} \right),
\end{equation}
always resulting in a pure quaternion, and reducing to a standard vector cross product if $u_0=v_0=0$.

\subsection{KS coordinates transformation}

\subsubsection{Generalized definition}

In a recent paper \citep{BL:17} we have proposed a generalized form of the standard KS transformation $\kappa$
that uses an arbitrary `defining vector' $\vv{c}$ with a unit norm and its respective
pure quaternion $\qq{c} = (0,\vv{c})$, so that
\begin{equation}\label{KSq}
 \kappa : \mathbb{H} \rightarrow \mathbb{H}' : \qq{v} \mapsto     \qq{x}    =  \frac{\qq{v} \, \qq{c} \, \overline{\qq{v}}}{\alpha},
\end{equation}
or, equivalently,
\begin{equation}
   \alpha \vv{x} = \left( v_0^2 - \vv{v}\bcd \vv{v} \right) \, \vv{c} + 2 \left(\vv{c}\bcd \vv{v}\right) \vv{v}
    + 2 v_0 \vv{v} \bcr \vv{c} =   \left(\vv{c} \bcd \vv{v}\right)\,\vv{v}+\left[\qq{v} \wedge (\qq{v} \wedge \qq{c})\right]^\natural, \label{KSgen}
\end{equation}
links the KS variables quaternion $\qq{v}$ with the original Cartesian coordinates $\vv{x} \in \mathbb{R}^3$,
the latter being the vector part of a pure quaternion
$\qq{x} = (0, \vv{x})$. A real, positive parameter
$\alpha$  was introduced by \citet{DEF}. They gave it the units of length, in order
to allow the KS  coordinates $v_j$  carry the same units as $x_j$. We adhere to this convention for a while,
although  other options will be presented in Section~\ref{KScan}.
With $|\qq{c}|=\|\vv{c}\|=1$, the KS transformation $\kappa$ admits the well known property
\begin{equation}\label{rgen}
   \| \vv{x} \| = r = \frac{\qq{v} \bcd \qq{v}}{\alpha}.
\end{equation}

\subsubsection{Fibres}

A non-injective nature of the KS map had been known since its origins, although only
recently it has been considered more an advantage than a nuisance \citep{RUP:16}.

Let us introduce a quaternion-valued function of angle $\phi$
\begin{equation}\label{fgen}
    \qq{q}(\phi) = (\cos{\phi}, \sin{\phi}\, \vv{c}),
\end{equation}
with a number of useful properties, like
\begin{eqnarray}
  |\qq{q}(\phi)| & = & 1, \\
  \qq{q}(\phi) \qq{q}(\psi) &=& \qq{q}(\phi+\psi), \\
 \left[\qq{q}(\phi)\right]^{-1} & = & \qq{q}(-\phi) = \overline{\qq{q}}(\phi), \\
 \qq{q}(\phi) \qq{c} \overline{\qq{q}}(\phi)& = &  \qq{c}, \label{prop:4}
\end{eqnarray}
and special values $\qq{q}(0) = \qq{e}_0$, $\qq{q}(\pi/2) = \qq{c}$.
The property (\ref{prop:4}) clearly implies that the KS transformation (\ref{KSq}) is only homomorphic: given some
representative KS quaternion $\qq{v}$, all
quaternions $\qq{v}\qq{q}(\phi)$ belonging to the fibre parameterized by $0 \leqslant \phi <  2 \pi$,
render the same vector $\vv{x}$, i.e. $\kappa(\qq{v}) = \kappa(\qq{v}\qq{q}(\phi))$. Indeed,
since (\ref{prop:4}) describes the rotation of vector $\vv{c}$ around the axis $\vv{c}$,
the left hand side of the equality can be substituted for $\qq{c}$ in eq.~(\ref{KSq}), and then
$\qq{v} \, \qq{c} \, \overline{\qq{v}} = (\qq{v} \qq{q}) \, \qq{c} \, \overline{(\qq{v} \qq{q})}$, leading to the same
$\qq{x}$.

On the other hand, one might ask about the possibility of generating the fibre through the left multiplication
by some quaternion function. Multiplying both sides of equality in (\ref{KSq}) by a quaternion $\qq{p}$
from the left and its conjugate from the right we find the condition
\[
  \qq{p} \qq{x} \overline{\qq{p}} = \frac{(\qq{p} \qq{v}) \, \qq{c} \, \overline{(\qq{p} \qq{v})} }{\alpha},
\]
where the left-hand side remains equal to $\qq{x} = (0,\vv{x})$ only if $\qq{p}$ is a function
\begin{equation}\label{fgen:x}
    \qq{p}(\phi) = ( \cos{\phi}, \sin{\phi} \hat{\vv{x}}),
\end{equation}
that rotates vector $\vv{x}$ around itself. Thus, given some representative KS quaternion $\qq{v}$, we can create the
fibre $\qq{p}(\phi)\qq{v}$ parameterized by $0 \leqslant \phi < 2 \pi$, such that $\kappa(\qq{v})=\kappa(\qq{p}(\phi)\qq{v}) = \qq{x}$.

The action of the fibre generators (\ref{fgen}) and (\ref{fgen:x}) with the same argument $\phi$ is equivalent; direct computation demonstrates that
\begin{equation}\label{equiv}
 \qq{p}(\phi)\qq{v} = \qq{v}\qq{q}(\phi), \qquad \mbox{ and }  \qquad \qq{p}(\phi)\qq{v}\overline{\qq{q}}(\phi) = \qq{v}.
\end{equation}
A kind of symmetry between the defining vector $\vv{c}$ and the normalized Cartesian position vector $\hat{\vv{x}}$ implied by
the form of $\qq{q}(\phi)$ and $\qq{p}(\phi)$ manifests also in the geometrical construction of the next section.

\subsection{KS quaternions \textit{more geometrico}}

\begin{figure}
	\includegraphics[width=\columnwidth]{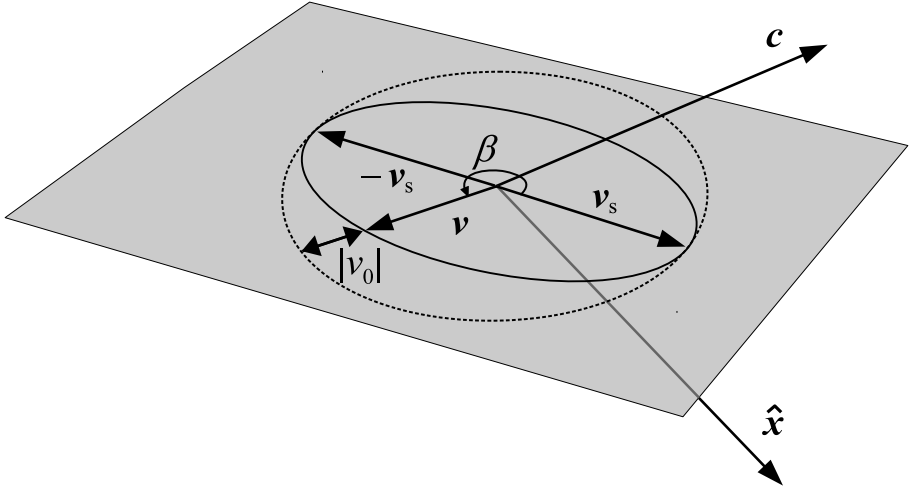}
    \caption{Geometrical construction for the vector part of the KS quaternion $\qq{v}$.}
    \label{fig:1}
\end{figure}

Given the transformation (\ref{KSgen}), let us polish the geometrical interpretation of the KS variables
proposed by \citet{Saha:09}. Scalar multiplication of both sides of (\ref{KSgen}) by $\qq{v}$ leads to the basic relation
\begin{equation}\label{cvcx}
    \hat{\vv{x}} \bcd \vv{v} = \vv{c} \bcd \vv{v},
\end{equation}
with two unit vectors $\vv{c}$ and $\hat{\vv{x}} = \vv{x}/r$. This property, valid for any scalar part $v_0$, means that
all quaternions $\qq{v}$ belonging to the fibre of given Cartesian vector $\vv{x}$, have vector parts $\vv{v} = \qq{v}^\natural$
forming the same angle with $\vv{c}$ and $\vv{x}$, hence lying in the symmetry plane of this pair of vectors. The plane, marked grey in Fig.~\ref{fig:1},
contains $\vv{c}+\hat{\vv{x}}$, and is perpendicular to $\vv{c} - \hat{\vv{x}}$. The norm $|\qq{v}|= \sqrt{r \alpha}$ is the upper bound on the length
of $\vv{v}$, so the dashed circle in Fig.~\ref{fig:1} has the radius $\sqrt{\alpha r}$.
Setting $v_0=0$ in equation~(\ref{KSgen}), we see that $\vv{x}$ is a linear combination of  $\vv{c}$ and $\vv{v}$, so the three vectors must be coplanar.
Accordingly, there are exactly two pure quaternions related with $\vv{x}$: $\qq{v}_\mathrm{s} = (0,\vv{v}_\mathrm{s})$, and $- \qq{v}_\mathrm{s}$, where
\begin{equation}\label{vs}
    \vv{v}_s = \sqrt{\alpha r} \frac{\vv{c} + \hat{\vv{x}}}{|| \vv{c} + \hat{\vv{x}}||},
\end{equation}
is the `Saha-Kustaanheimo-Stiefel (SKS)vector' of \citet{BL:17}.

The entire fibre $\qq{v}$ can be generated from $\qq{v}_\mathrm{s}$ by the application of
the generator (\ref{fgen}),
\begin{equation}\label{fibs}
    \qq{v} = \qq{v}_\mathrm{s} \,\qq{q}(-\phi),
\end{equation}
leading to
\begin{eqnarray}
% \nonumber to remove numbering (before each equation)
  v_0 &=&   \vv{c} \bcd \vv{v}_\mathrm{s} \sin{\phi}, \\
  \vv{v} &=& \cos{\phi} \vv{v}_\mathrm{s} + \sin{\phi} \left(\vv{c} \, \bcr \, \vv{v}_\mathrm{s}  \right).
\end{eqnarray}
The latter of the formulae is a parametric equation of an ellipse with the major semi-axis $\sqrt{\alpha r}$
and the eccentricity $\sqrt{(1+\vv{c} \bcd \hat{\vv{x}})/2}$. The ellipse is drawn with a solid line in Fig.~\ref{fig:1}.
The position angle $\beta$ in the figure should not be confused with the parametric longitude $\phi$;
the angles are related by the   formula
\begin{equation}\label{tnb}
    \tan \beta = \sqrt{\frac{1 -  \vv{c} \bcd \hat{\vv{x}} }{2}} \,  \tan \phi.
\end{equation}
The line segment with arrowheads at both ends in Fig.~\ref{fig:1}, complements the length of $\vv{v}$ to the full
value $\sqrt{\alpha r}$, so its length can be interpreted as the absolute value of the scalar part of $\qq{v}$.

Of course, the generic picture shown in Fig.~\ref{fig:1} does not include the special case of the parallel $\vv{x}$ and $\vv{c}$.
If $\hat{\vv{x}} = \vv{c}$, the fibre degenerates to the set of quaternions having vector part aligned with $\vv{c}$, i.e.
\begin{equation}\label{cisx}
    \qq{v} = \sqrt{\alpha r} \left( \sin{\phi} , \cos{\phi} \vv{c} \right),
\end{equation}
with $\qq{v}_\mathrm{s} = (0,\sqrt{\alpha r}\, \vv{c})$. The eccentricity of the ellipse from Fig.~\ref{fig:1} attains the value 1, so
the ellipse degenerates into a straight segment. The shaded plane from the figure is no longer defined.

But if $\hat{\vv{x}} = -\vv{c}$, the situation is different. Observing that then the ellipse from Fig.~\ref{fig:1} turns into a circle,
we conclude that the fibre consists exclusively of the pure quaternions $\qq{v} = (0, \sqrt{\alpha r} \,\hat{\vv{f}})$,
where $\hat{\vv{f}}$ is any vector orthogonal to $\vv{c}$.

\subsection{Bilinear form $\mathcal{J}$ and LC planes}
\label{JLC}

\subsubsection{Definitions}

The skew-symmetric bilinear form $\mathcal{J} : \mathbb{H} \times \mathbb{H} \rightarrow \mathbb{R}$,
introduced by \citet{Kust:64} and discussed in later works, can be
generalized to an arbitrary defining vector $\vv{c}$ as
\begin{equation}\label{Jdef}
    \mathcal{J}(\qq{v},\qq{w}) = \left(  \bar{\qq{v}}  \wedge  \bar{\qq{w}}  \right) \cdot \qq{c}
    = - v_0 \vv{w} \bcd \vv{c} + w_0 \vv{v} \bcd \vv{c} +  \left( \vv{v} \bcr \vv{w} \right) \bcd \vv{c}.
\end{equation}
The form plays a central role in the KS formulation of motion. If the motion can be restricted to
the linear subspace of $\mathbb{H}$ spun by two basis quaternions $\qq{u}$ and $\qq{w}$, such that
$\mathcal{J}(\qq{u},\qq{w})=0$, the KS transformation reduces to the Levi-Civita transformation \citep{LC:1906}.
For this reason, a two-dimensional subspace $P$ of quaternions
being the linear combinations of $\qq{u}$ and $\qq{w}$, hence such that the form $\mathcal{J}$ on any two
of them equals 0, was dubbed the `Levi-Civita plane' by \citet{StS:71}. We will use the name `LC plane',
although, strictly speaking, a (hyper-)plane in a space of dimension 4 should be spun by 3 basis quaternions.

Repeating the proof of Theorem 3 from \citet{DEF} in our generalized framework, we conclude that for any unitary quaternion $\qq{u}$
selected for the orthonormal basis of LC plane $P$, the second basis quaternion should be
\begin{equation}\label{2base}
    \qq{w} = \qq{u} (0, \vv{f}) = ( - \vv{u} \cdot \vv{f}, u_0 \vv{f} + \vv{u} \times \vv{f}),
\end{equation}
where $\vv{f}$ is any unitary vector orthogonal to the defining vector, i.e. $\vv{f} \cdot \vv{c} =0$, and $||\vv{f}||=1$.
This meaning of the symbol $\vv{f}$ will be held throughout the text.
The basis is indeed orthonormal, since $\qq{u}\cdot \qq{w} = 0$, and $|\qq{u}|=|\qq{w}|=1$, by the definition of $\qq{u}$ and $\vv{f}$.

\subsubsection{KS map of an LC plane}
\label{KSLC}

Once the LC plane has been defined, a question arises about the possibility of restricting the motion in KS variables to this subspace.
But such restriction implies that the motion in `physical' configuration space $\mathbb{R}^3$ is planar.

Let us prove that KS transformation maps any quaternion in the LC plane $P$ onto a plane $\Pi$ in `physical' $\mathbb{R}^3$ space.
Using the basis of two orthonormal quaternions $\qq{u}$ and $\qq{w} = \qq{u} (0, \vv{f})$, we consider
their linear combination
\begin{equation}\label{linco}
    \qq{v} = \xi \qq{u} + \eta \qq{w} = \qq{u}\,(\xi,\,\eta \vv{f}),
\end{equation}
with real parameters $\xi,\eta$ having the dimension of length. The KS transform of these $\qq{v}$, belonging to $P$,
is,by the definition (\ref{KSq}),
\begin{equation} \label{trav}
  \qq{x} = \kappa{(\qq{v})} = \frac{\qq{u}\,(\xi,\,\eta \vv{f})\,\qq{c}\,(\xi,\,-\eta \vv{f})\,\overline{\qq{u}}}{\alpha}.
\end{equation}
Thanks to the orthogonality of $\vv{c}$ and $\vv{f}$, the product in the middle evaluates to
\begin{equation}\label{mid}
    (\xi,\,\eta \vv{f})\,\qq{c}\,(\xi,\,-\eta \vv{f}) = \left(\xi^2 - \eta^2\right) ( 0, \vv{c}) + 2 \xi \eta ( 0, \vv{f} \times \vv{c}),
\end{equation}
so the vector part of $\kappa{(\qq{v})}$ is a linear combination of two fixed, orthonormal vectors
\begin{equation}\label{trax}
  \vv{x} = \frac{ \xi^2 - \eta^2 }{\alpha}\, \hat{\vv{x}}_1 + \frac{2 \xi \eta }{\alpha}\,\hat{\vv{x}}_2,
\end{equation}
where
\begin{eqnarray}
  \hat{\vv{x}}_1 &=&  \left[ \qq{u} (0,\vv{c}) \overline{\qq{u}} \right]^\natural  = (2 u_0^2-1) \vv{c} + 2 (\vv{u}\cdot \vv{c}) \, \vv{u}
  + 2 u_0 \vv{u} \times \vv{c}, \label{x1}\\
  \hat{\vv{x}}_2 &=&  \left[ \qq{u} (0,\vv{f} \times \vv{c}) \overline{\qq{u}} \right]^\natural \nonumber \\
  & = & (2 u_0^2-1) (\vv{f} \times \vv{c}) + 2 (\vv{u}\cdot (\vv{f} \times \vv{c})) \, \vv{u}  + 2 u_0 \vv{u} \times (\vv{f} \times \vv{c}). \label{x2}
\end{eqnarray}
Since eq.~(\ref{trax}) is actually a parametric equation of a plane in $\mathbb{R}^3$, we have
demonstrated that the KS transformation of any LC plane $P \subset \mathbb{H}$ is a
plane $\Pi$ in $\mathbb{R}^3$ (or in $\mathbb{H}'$, depending on the context).
The parameters $\xi$, $\eta$ become parabolic coordinates on $\Pi$, i.e. the usual Levi-Civita variables.

The vector normal to the plane $\hat{\vv{x}}_3 = \hat{\vv{x}}_1 \times \hat{\vv{x}}_2$, can be most easily
derived in terms of the quaternion cross product (\ref{qop}), with the first lines of (\ref{x1}) and (\ref{x2})
substituted. Thus, letting $\qq{x}_j = (0,\hat{\vv{x}}_j)$, and $\qq{b} = (0, \vv{f} \times \vv{c})$,
we find
\begin{eqnarray}
  \qq{x}_3  & = &   \frac{1}{2} \left[ \overline{\qq{x}}_2 \qq{x}_1
 - \overline{\qq{x}}_1 \qq{x}_2 \right] \nonumber \\
 & = & \frac{1}{2} \left[ ( \qq{u} \qq{b} \overline{\qq{u}})( \qq{u} \overline{\qq{c}} \,\overline{\qq{u}})
  -   ( \qq{u} \overline{\qq{c}} \, \overline{\qq{u}})( \qq{u} \qq{b} \overline{\qq{u}})\right]  \nonumber \\
 &=& \qq{u} (\qq{c} \wedge \qq{b} ) \overline{\qq{u}}  =   \qq{u}\,(0,\vv{f})\, \overline{\qq{u}}  \nonumber \\
 &=& \left( 0, (2 u_0^2 -1) \vv{f} + 2 ( \vv{u} \cdot \vv{f} ) \vv{u} + 2 u_0 \vv{u} \times \vv{f}\right). \label{x3}
\end{eqnarray}
Thanks to the above equation, we can relate the choice of the LC plane basis to the orientation of $\Pi$.
The cosine of the angle between the defining vector $\vv{c}$ and the normal to the plane of motion $\hat{\vv{x}}_3$ is
given by the scalar product
\begin{equation}\label{cx3}
    \vv{c} \cdot \hat{\vv{x}}_3 = 2 (\vv{u} \cdot \vv{f})  (\vv{u} \cdot \vv{c}) - 2 u_0 \vv{u} \cdot (\vv{c} \times \vv{f}).
\end{equation}

\subsubsection{KS1 and KS3 setup}

Some particular choices of the first basis quaternion $\qq{u}$ deserve a special comment.
Inspecting eq.~(\ref{cx3}), we notice three obvious cases leading to $\vv{c}$ positioned in the plane of motion:
a pure scalar $\qq{u} = \pm (1,\vv{0})$, or pure quaternions: $\qq{u} = (0, \pm \vv{f})$, and $\qq{u} = (0, \pm \vv{c})$.
The basis vectors $\hat{\vv{x}}_1$, resulting  from eq.~(\ref{x1}), are $\vv{c}$, $- \vv{c}$, and $\vv{c}$, respectively.
The last case, i.e. $\qq{u} = (0, \vv{c})$, has been the most common choice in celestial mechanics since the first paper of \citet{Kust:64}.
It allows the most direct identification of the LC plane with the plane of motion, both spanned by the same
vectors (or pure quaternions)
$\qq{u}^\natural = \hat{\vv{x}}_1 = \vv{c}$, and $\qq{w}^\natural = \hat{\vv{x}}_2 = \vv{c} \times \vv{f}$.
The freedom of choice for $\vv{f}$ (any vector perpendicular to $\vv{c}$) permits to identify $\vv{c}$ and $-\vv{f}$ with
the basis vector $\vv{e}_1$ and $\vv{e}_3$ of the particular reference frame used to describe the planar ($x_3=0$) motion.
For this reason, let us call the KS transformation based upon the paradigmatic choice $\vv{c}=\vv{e}_1$, the KS1 transformation.

Remaining in the domain of pure quaternions, let us consider $\qq{u} = (0,\vv{u})$. Without loss of generality,
we can assume $\vv{u} = \cos{\psi} \vv{c} + \sin{\psi} \vv{f}$, with $0 \leqslant \psi \leqslant \pi$.
Then, according to eq.~(\ref{cx3}), we have $\vv{c} \cdot \hat{\vv{x}}_3 = \sin{2 \psi}$, so an appropriate choice of the
parameter $\psi$ may lead to any orientation of the orbital plane with respect to $\vv{c}$.
In particular, the defining vector will coincide with $\hat{\vv{x}}_3$ when $\psi=\pi/2$. The LC plane
spanned by the basis quaternions
\begin{equation}\label{bas3}
    \qq{u} = \left( 0 , \frac{\vv{c}+\vv{f}}{\sqrt{2}} \right), \qquad \qq{w} = \left( - \frac{1}{\sqrt{2}}, \frac{\vv{c} \times \vv{f}}{\sqrt{2}} \right),
\end{equation}
is mapped onto the plane of motion $\Pi$ with basis vectors
$\hat{\vv{x}}_1 = \vv{f}$, and $\hat{\vv{x}}_2 = \vv{c} \times \vv{f}$ -- both orthogonal to $\vv{c}$.
Thus the choice of  $\vv{c}=\vv{e}_3$, and $\vv{f}=\vv{e}_1$ leads to the KS3 transformation, which may look less attractive than
KS1, with its LC plane no longer consisting of pure quaternions. Indeed, it is not practiced in celestial mechanics,
save for two exceptions known to the authors \citep{Saha:09,BL:17}. In physics, however,
the KS3 transformation is common at least since 1970's \citep[e.g.][]{DuKl:79,Cordani,Diaz:10,Egea:11,vdM:16};
there are good reasons for this, but they come out only in the context of dynamics and symmetries
of a perturbed Kepler (or Coulomb) problem.

\section{Canonical KS variables in the extended phase space}
\label{KScan}

In contrast to earlier works, let us consider  from the onset a canonical problem  in the extended phase space
$(x^\ast,\vv{x},X^\ast,\vv{X})$, with a Hamiltonian
\begin{equation}\label{Ham}
    \cH(x^\ast,\vv{x},X^\ast,\vv{X}) = \cH_0(\vv{x},\vv{X}) +  \mathcal{R}(x^\ast,\vv{x},\vv{X})+X^\ast = 0,
\end{equation}
where the Keplerian term
\begin{equation}\label{Hkep}
    \cH_0 = \frac{\vv{X} \cdot \vv{X}}{2} - \frac{\mu}{r},
\end{equation}
depends on the Cartesian coordinates $\vv{x}$, their conjugate momenta $\vv{X}$ and the gravitational parameter $\mu$.
The time-dependent perturbation $\mathcal{R}(t,\vv{x},\vv{X})$ is converted into a conservative term by substituting a formal, time-like
coordinate $x^\ast$ for physical time $t$. The fact that $x^\ast(t)=t$ is a direct consequence of the way its conjugate momentum $X^\ast$ appears in
equation (\ref{Ham}), because
\begin{equation}\label{sdef}
    \dot{x^\ast} = \frac{\partial \cH}{\partial X^\ast} = 1,
\end{equation}
and an appropriate choice of the arbitrary constant leads to the identity map of $t$ on $x^\ast$.
The momentum $X^\ast$ itself evolves according to
\begin{equation}\label{Sdef}
    \dot{X}^\ast = - \frac{\partial \cH}{\partial x^\ast} = - \frac{\partial \mathcal{R}}{\partial x^\ast},
\end{equation}
counterbalancing the variations of energy in nonconservative problems, or staying constant in the conservative case.

If the same problem is to be handled canonically in terms of the KS coordinates,
their conjugate momenta $\qq{V}$ are implicitly defined through
\begin{equation}\label{VtoXq}
    \qq{X} = \frac{\qq{V} \qq{c} \bar{\qq{v}}}{2r}, \mbox{~~~~or~~~~}\qq{V} = \frac{2\,\qq{X} \,\qq{v} \,\bar{\qq{c}}}{\alpha}.
\end{equation}
In this transformation we postulate
\begin{equation}\label{JvV}
    \mathcal{J}(\qq{v},\qq{V}) = \left(  \bar{\qq{v}}  \wedge  \bar{\qq{V}}  \right) \cdot \qq{c} = 0,
\end{equation}
to secure
\begin{equation}\label{X0}
    X_0 = \frac{\mathcal{J}(\qq{v},\qq{V})}{2 r} =0,
\end{equation}
so that $\qq{X} = (0, \vv{X})$ remains a pure quaternion.
The transformation $\mathbb{R}^2 \times \mathbb{H}^2 \rightarrow \mathbb{R}^2 \times \mathbb{H}' \times \mathbb{H}'$, which maps $(v^\ast,\qq{v},V^\ast,\qq{V})
 \mapsto (x^\ast,\vv{x},X^\ast,\vv{X})$ according
to the definitions (\ref{KSq}), (\ref{VtoXq}) and the identities $x^\ast = v^\ast$, $X^\ast = V^\ast$,
is known to be weakly canonical (i.e. canonical only on a specific manifold (\ref{JvV})).

Let us now generalize the transformation by allowing that $\alpha$, instead of being a fixed parameter, is an arbitrary differentiable function
of the energy-like momentum $X^\ast$ or $V^\ast$. A similar assumption was recently made for the Levi-Civita transformation \citep{BL:18a}. The necessity, or at least usefulness of such a generalization will not be clear until the action-angle variables are introduced,
but it has to be introduced already at this stage. If the generalized transformation is to be kept weakly canonical, while maintaining
the direct relation $V^\ast = X^\ast$,  the new formal time-like variable
$v^\ast$ should differ from $x^\ast$.\footnote{Giving credit to previous applications of 
this idea in \citet{BL:18a}, we have overlooked \citet{StS:71}, much earlier than \citet{Zhao:16}.}
Then, the transformation
\begin{equation}\label{lamb}
    \lambda : \quad (v^\ast,\qq{v},V^\ast,\qq{V}) \mapsto (x^\ast,\vv{x},X^\ast,\vv{X}),
\end{equation}
conserves the Pfaffian one-form up to the total differential of a primitive function $Q$ \citep{AKN:97}
\begin{equation}\label{difform}
 V^\ast\,\mathrm{d}v^\ast + \qq{V} \cdot \mathrm{d}\qq{v}- X^\ast\,\mathrm{d}x^\ast - \vv{X} \cdot \mathrm{d}\vv{x}   = \mathrm{d}Q
 + \frac{\cJ(\qq{v},\qq{V}) \cJ(\qq{v} ,\mathrm{d} \qq{v})}{\qq{v}\cdot \qq{v}},
\end{equation}
provided
\begin{eqnarray}
  Q &=&  \left[ \frac{X^\ast}{\alpha} \frac{\partial \alpha}{\partial X^\ast}\right]\, \vv{x} \cdot \vv{X}
  = \left[\frac{V^\ast}{ \alpha} \frac{\partial \alpha}{\partial V^\ast}\right]\, \frac{\qq{v} \cdot \qq{V}}{2}, \\
  x^\ast &=& v^\ast - \frac{Q}{V^\ast}, \label{xsvs}
\end{eqnarray}
and with a necessary condition of $\cJ(\qq{v},\qq{V})=0$.

It is worth noting, that with an elementary choice of $\alpha = k_1 (X^\ast)^{k_2}$, the expression in the square bracket evaluates
to a single number $k_2$, and the multiplier $k_1$ has no influence on canonicity, hence it can be selected at will -- for example to conserve
(or to modify) the units of time and length.

In order to convert the Hamiltonian (\ref{Ham}) into a perturbed harmonic oscillator, the independent variable has to be changed from the physical time
$t$ to the Sundmann time $\tau$, related by
\begin{equation}\label{sund}
    \frac{\mathrm{d}\tau}{\mathrm{d} t} =  \frac{\alpha}{4 r} = \frac{\alpha^2}{4 \,\qq{v} \cdot \qq{v}},
\end{equation}
involving $\alpha$ as a function of $V^\ast$ or $X^\ast$.
Transforming the Hamiltonian (\ref{Ham}) by the composition of $\lambda$ and $t \mapsto \tau$, we obtain
\begin{eqnarray}\label{K}
    \cK(v^\ast,\qq{v},V^\ast,\qq{V}) & = & \cK_0(\qq{v},V^\ast,\qq{V})+ \mathcal{P}(v^\ast,\qq{v},V^\ast,\qq{V}) = 0, \\
    \cK_0(\qq{v},V^\ast,\qq{V}) & = & \frac{\qq{V} \cdot \qq{V}}{2} + \frac{\omega^2\, \qq{v} \cdot \qq{v}}{2} - \frac{4\mu}{\alpha}
    + \frac{\alpha \mathcal{J}(\qq{v},\qq{V})^2}{2 \, |\qq{v}|^2},~~~~ \label{K0} \\
    \mathcal{P}(v^\ast,\qq{v},V^\ast,\qq{V}) & = & \frac{4 r}{\alpha} \mathcal{R}^\star(v^\ast,\qq{v},V^\ast,\qq{V}),
\end{eqnarray}
where $\mathcal{R}^\star$ is the perturbation Hamiltonian $\mathcal{R}(x^\ast,\vv{x},\vv{X})$ expressed in terms of the extended KS coordinates and momenta,
and
\begin{equation}\label{om0}
    \omega = \frac{2 \sqrt{2 V^\ast}}{\alpha},
\end{equation}
will have a constant value only if the original Hamiltonian $\cH$ does not depend on time. Let us emphasize, that now every function of
$\vv{x}$, when expressed in terms of $\qq{v}$, will generally depend on the energy-like momentum $V^\ast$ as well, due to its presence in $\alpha$.
Noteworthy, the simplification to $\omega=1$ can be achieved by assuming $\alpha = \sqrt{8 V^\ast}$, which makes the Sundmann time dimensionless.
Choosing $\alpha = \mu/ V^\ast$, is roughly equivalent to $\alpha = 2 a$, in terms of the Keplerian orbit semi-axis $a$.

\section{Action-angle variables}
\label{AAs}

\subsection{LLC and LCF variables}

When the motion is planar, with $x_3=0$, an appropriate action-angle set $l,g,L,G$ can be created using a combination of the Levi-Civita
\citep{LC:1906} and Lissajous transformations \citep{DepWil:91}. This approach has been recently revisited and discussed
by \citet{BL:18a}. Viewed as the special case of the KS framework, the Lissajous-Levi-Civita (LLC)
variables are inherently attached to the KS1 setup, requiring the identification of
the LC plane $P \subset \mathbb{H}'$ of pure quaternions and the plane of motion $\Pi \subset \mathbb{R}^3$.
A generalization of this approach was proposed by \citet{Zhao:15}. Roughly speaking, he attached the LC plane to an osculating plane of motion $\Pi$,
and added the third action-angle pair $h,H$ orienting $\Pi$ in $\mathbb{R}^3$ by direct analogy with the third Delaunay pair: longitude of the ascending node,
and projection of the angular momentum on the axis $\hat{\vv{x}}_3$. As noted by the author, this approach has the same drawbacks
as in the Dalunay set -- in particular, the singularity when the orbit in physical space is rectilinear, thus having no unique orbital plane.

\subsection{Lissajous-Kustanheimo-Stiefel (LKS) variables}

\subsubsection{Intermediate set}

The starting point for the new set of variables we would like to propose is completely different than in \citet{Zhao:15}.
First, we choose the KS3 framework, assuming the defining vector $\vv{c}=\vv{e}_3$. Then, we select two subspaces of $\mathbb{H}$:
$P_{03}$ with the basis $\qq{e}_0, \qq{e_3}$, and $P_{12}$ spanned by $\qq{e}_1$ and $\qq{e}_2$. None of them is a Levi-Civita plane,
because in the KS3 framework
\begin{eqnarray}
\cJ(\qq{e}_0,\qq{e}_3) & = &  \cJ((1,\vv{0}),(0,\vv{c})) = -1, \nonumber \\
\cJ(\qq{e}_0,\qq{e}_3) & = &  \cJ((1,\vv{f}),(0,\vv{c} \times \vv{f}))=1.
\end{eqnarray}
Thus, even for the planar case, we do not restrict motion to an invariant plane $P$, but merely project $\qq{v}$ on two orthogonal subspaces.
The orthogonality is readily checked by
\begin{equation}\label{orto}
    (v_0 \qq{e}_0 + v_3 \qq{e}_3) \cdot (v_1 \qq{e}_1 + v_2 \qq{e}_2) = 0.
\end{equation}
On each plane, with $(i,j) = (0,3)$, or $(i,j) = (1,2)$, we perform the Lissajous transformation of \citet{Dep:91}
\begin{eqnarray}
    v_i & = & \sqrt{\frac{L_{ij}+G_{ij}}{2 \omega}} \cos{(l_{ij}+g_{ij})} - \sqrt{\frac{L_{ij}-G_{ij}}{2 \omega}} \cos{(l_{ij}-g_{ij})}, \label{LT1} \\
    v_j & = & \sqrt{\frac{L_{ij}+G_{ij}}{2 \omega}} \sin{(l_{ij}+g_{ij})} + \sqrt{\frac{L_{ij}-G_{ij}}{2 \omega}} \sin{(l_{ij}-g_{ij})}, \label{LT2} \\
    V_i & = & -\sqrt{\frac{\omega (L_{ij}+G_{ij})}{2}} \sin{(l_{ij}+g_{ij})}  + \sqrt{\frac{\omega (L_{ij}-G_{ij})}{2}} \sin{(l_{ij}-g_{ij})}, \label{LT3} \\
    V_j & = & \sqrt{\frac{\omega (L_{ij}+G_{ij})}{2}} \cos{(l_{ij}+g_{ij})}  + \sqrt{\frac{\omega (L_{ij}-G_{ij})}{2}} \cos{(l_{ij}-g_{ij})}. \label{LT4}
\end{eqnarray}
Similarly to \citet{BL:18a}, we allow $\omega > 0$ to be a function of $V^\ast$, as given by equation (\ref{om0}) -- both directly, and through $\alpha$.
This requires a new time-like variable $s$ to be different from $v^\ast$, while retaining its conjugate $S=V^\ast$. Only then, the
1-forms are conserved up to the total differential
\begin{equation}\label{pfaf:01}
    L_{03} \mathrm{d}l_{03} +   G_{03} \mathrm{d}g_{03} +  L_{12} \mathrm{d}l_{12} +   G_{12} \mathrm{d}g_{12}
    + S \mathrm{d}s -  \qq{V} \cdot \mathrm{d} \qq{v} - V^\ast \mathrm{d} v^\ast = \mathrm{d}Q^\ast,
\end{equation}
with
\begin{equation}\label{Qast}
    Q^\ast =  -\frac{\qq{v} \cdot \qq{V}}{2} \left( 1 - \frac{ S}{\omega}\frac{\mathrm{d} \omega}{\mathrm{d} S} \right),
\end{equation}
and
\begin{equation}\label{vss}
    v^\ast = s  -\frac{\qq{v} \cdot \qq{V}}{2 \omega}  \frac{\mathrm{d} \omega}{\mathrm{d} S},
\end{equation}
where
\begin{equation}\label{Vvi}
 \qq{v} \cdot \qq{V} = \sqrt{L_{03}^2-G_{03}^2} \sin{2 l_{03}} + \sqrt{L_{12}^2-G_{12}^2} \sin{2 l_{12}}.
\end{equation}

The Hamiltonian function (\ref{K}) is converted into the sum of
\begin{equation}
\cK'_0   =   \omega L_{03} + \omega L_{12}   - \frac{4\mu}{\alpha}
    + \frac{\alpha \left(G_{03}-G_{12}\right)^2}{8 \, |\qq{v}|^2},
\end{equation}
and of the perturbation $\mathcal{P}$ expressed in terms of the Lissajous variables.

\begin{figure}
	\includegraphics[width=\columnwidth]{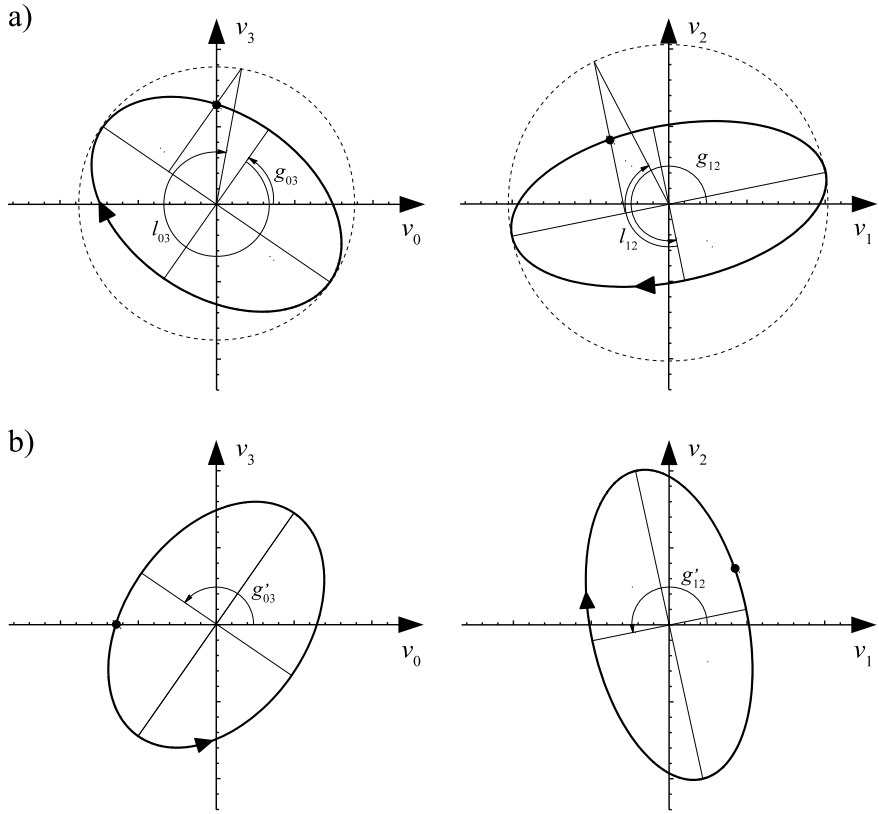}
    \caption{The motion in two configuration planes of the KS3 variables for the Kepler problem. a) Initial
    conditions $\qq{v}$ set according to eq.~(\ref{vs}). b) Initial conditions
    are multiplied by $\qq{q}(\pi/2)$. More details in the text. }
    \label{fig:2}
\end{figure}
This transformation is merely an intermediate step, but before the final move let us inspect
the meaning and properties of the variables in the Kepler problem defined by $\cK'_0=0$.
As a generic example we take a heliocentric orbit in physical phase space with the following Keplerian elements:
major semi-axis $a=10 \mathrm{au}$, eccentricity $e=0.5$,
inclination $I=10^\circ$, argument of perihelion $\omega_\mathrm{o} = 60^\circ$, longitude of the ascending node $\Omega=10^\circ$,
and the initial true anomaly $f=60^\circ$. From these elements we compute first the position $\vv{x}(0)$ and momentum $\vv{X}(0)$,
and then the representative KS3 quaternions $\qq{v}(0)$ and $\qq{V}(0)$ -- an SKS vector given by equation (\ref{vs}), and its conjugate momentum
defined by equation (\ref{VtoXq}), both with $\qq{c}=\qq{e}_3$. These initial conditions are labeled with black dots in Fig.~\ref{fig:2}a.
The ellipses described in the $(v_0,v_3)$ and $(v_1,v_2)$ planes have different semi-axes and different eccentricities; however, both are traversed in the
same direction -- retrograde (clockwise) in the discussed example. The retrograde motion follows from the fact that $G_{03}=G_{12} < 0$ (the momenta are
equal due to the postulate (\ref{JvV}), where $\left(  \bar{\qq{v}}  \wedge  \bar{\qq{V}}  \right) \cdot \qq{e}_3 = (G_{03}-G_{12})/2$).
The constant angles $g_{03}$ and $g_{12}$, measured counterclockwise, position the ellipses in the coordinate planes. The initial
angles $l_{03}$ and $l_{12}$ are marked according to the geometrical construction similar to that of the eccentric anomaly. Comparing our Fig.~\ref{fig:2}
with Fig.~1 of \citet{Dep:91}, the readers may note the reverse direction of the $l_{ij}$ angle. The difference comes from the fact that \citet{Dep:91}
assumed $G > 0$, i.e. the prograde (counterclockwise) motion along the Lissajous ellipse. Yet, regardless of the sign of $G_{ij}$, equations of motion imply
$\rd l_{03}/\rd \tau = \rd l_{12}/\rd \tau = \omega > 0$.

Each Lissajous ellipse has the major semi-axis $a_{ij}$ and the minor semi-axis $b_{ij}$ defined by the two momenta and frequency
\begin{equation}\label{abij}
    a_{ij} = \frac{\sqrt{L_{ij}+G_{ij}}+\sqrt{L_{ij}-G_{ij}}}{2 \omega}, \quad  b_{ij} = \frac{|\sqrt{L_{ij}+G_{ij}}-\sqrt{L_{ij}-G_{ij}}|}{2 \omega}.
\end{equation}
The absolute value operator is necessary for $G_{ij}<0$, unless one adopts a convention of negative minor semi-axis for an ellipse traversed clockwise.

Another point worth observing is the ambiguity in the choice of the $l_{ij}$ and $g_{ij}$ pair. Their values are determined from two possible sets of four equations
linking quadratic forms of $v_i,v_j,V_i,V_j$ with sine and cosine functions of the angles. Regardless of whether we use
\[\sin{2 g_{ij}}, \cos{2 g_{ij}}, \sin{2 l_{ij}}, \cos{2 l_{ij}}, \]
or
\[ \sin{(l_{ij}+ g_{ij})}, \cos{(l_{ij}+g_{ij})},
\sin{(l_{ij}-g_{ij})}, \cos{(l_{ij}-g_{ij})},\]
the solution will always result in two pairs: $(l_{ij},g_{ij})$, and $(l_{ij}+\pi, g_{ij}+\pi)$ -- both giving the same
values of the sine and cosine.\footnote{The statements about `the Lissajous variables [...] determined unambiguously from the Cartesian variables' made by \citet{Dep:91}
should not be taken too literally.} In other words, one of the two minor semi-axes in each of the ellipses in Fig.~\ref{fig:2} can be chosen at will
as the reference one.

Recalling the fibration property of the KS variables, we have plotted the ellipses obtained from the same Cartesian $\vv{x}(0)$ and $\vv{X}(0)$, but with the
KS3 initial conditions $\qq{v}(0)$ and $\qq{V}(0)$ right-multiplied by $\qq{q}(\pi/2)=\qq{c}=\qq{e}_3$, according to equation~(\ref{fgen}) in the KS3 case.
The results are displayed in Fig.~\ref{fig:2}b. Not only the initial conditions, abut the entire ellipses are rotated by $90^\circ$ in the $(v_0,v_3)$ plane,
and by $-90^\circ$ in the $(v_1,v_2)$ plane. The momenta $L_{ij}, G_{ij}$, and the angles $l_{ij}$ remain intact, compared to Fig.~\ref{fig:2}a. The new
angles positioning the ellipses are $g'_{03} = g_{03}+\pi/2$, and $g'_{12} = g_{12}-\pi/2$, but their sum has not changed: $g'_{03}+g'_{12} = g_{03}+g_{12}.$

\subsubsection{Final transformation}

Bearing in mind the example shown in Fig.~\ref{fig:2}, we can establish the final set of the LKS variables by defining four action-angle pairs
\begin{eqnarray}
l & = & \frac{1}{2} \left(l_{12}+l_{03}\right), \nonumber \\
\lambda & = & \frac{1}{2} \left(l_{12}-l_{03}\right), \nonumber \\
g & = & \frac{1}{2} \left(g_{12}+g_{03}\right), \nonumber \\
\gamma & = & \frac{1}{2} \left(g_{12}-g_{03}\right),  \label{LKS:1}\\
 L & = &  L_{12}+L_{03}, \nonumber \\
 \Lambda & = & L_{12} - L_{03}, \nonumber \\
 G & = & G_{12} + G_{03}, \nonumber \\
 \Gamma & = & G_{12}-G_{03}, \nonumber
\end{eqnarray}
with $s$ and $S$ retained unaffected. One may easily verify that (\ref{LKS:1}) amounts to an elementary Mathieu transformation, thus the complete
composition
\[
\zeta:~(x^\ast,\qq{x},X^\ast,\qq{X}; t) \rightarrow (s,l,\lambda,g,\gamma,S,L,\Lambda,G,\Gamma; \tau),
\]
is a weakly canonical, dimension raising transformation.
The Hamiltonian $\cH$ from equation~(\ref{Ham}) is transformed into
\begin{equation}\label{M}
     \cM(s,l,\lambda,g,S,L,\Lambda,G,\Gamma) = \cM_0(l,\lambda,S,L,\Lambda,\Gamma) + \mathcal{Q}(s,l,\lambda,g,S,L,\Lambda,G,\Gamma) = 0,
\end{equation}
where
\begin{equation}\label{M0}
    \cM_0 = \omega(S)\,L - \frac{4 \mu}{\alpha(S)} + \frac{\Gamma^2}{8 r},
\end{equation}
and $\mathcal{Q}$ is the pullback of $\frac{4 r}{\alpha(S)} \mathcal{R}(x^\ast,\vv{x},\vv{X})$ by $\zeta$.

Expressing the Cartesian variables from the initial extended phase space in terms of the LKS variables, we
first introduce six actions-dependent coefficients
\begin{eqnarray}
 A_1 &=& \frac{1}{2} \sqrt{(L + G)^2 - (\Lambda + \Gamma)^2}, \nonumber  \\
 A_2 &=& \frac{1}{2} \sqrt{(L - G)^2 - (\Lambda - \Gamma)^2}, \nonumber  \\
 B_1 &=& \frac{1}{2} \sqrt{(L + \Lambda)^2 - (G + \Gamma)^2}, \nonumber \\
 B_2 &=& \frac{1}{2} \sqrt{(L - \Lambda)^2 - (G - \Gamma)^2}, \label{coefs} \\
 C_1 &=& \frac{1}{2} \sqrt{(L + \Gamma)^2 - (G + \Lambda)^2}, \nonumber \\
 C_2 &=& \frac{1}{2} \sqrt{(L - \Gamma)^2 - (G - \Lambda)^2}, \nonumber
\end{eqnarray}
allowing a compact formulation of the expressions for coordinates
\begin{eqnarray}
x_0 & = & 0, \\
  x_1 &=&  \frac{1}{  \sqrt{8 S}} \left(   A_1 \sin{2 (l + g)} - A_2 \sin{2 (l - g)} \right.  \nonumber \\
   & & \left.  - C_1 \sin{2 (g + \lambda)} - C_2  \sin{2 (g - \lambda)}   \right),  \label{xLKS}\\
  x_2 &=&  \frac{1}{  \sqrt{8 S}} \left( - A_1 \cos{2 (l + g)}  - A_2  \cos{2 (l - g)} \right.  \nonumber \\
   & & \left. + C_1  \cos{2 (g + \lambda)} + C_2 \cos{2 (g - \lambda)} \right), \\
  x_3 &=&  \frac{1}{ \sqrt{8 S}} \left( -  \Lambda + B_1 \cos{2 (l + \lambda)}  - B_2 \cos{2 (l - \lambda)}  \right),
\end{eqnarray}
and momenta
\begin{eqnarray}
% \nonumber to remove numbering (before each equation)
  X_0 &=&   \frac{\Gamma}{2 r} = 0,  \nonumber \\
  X_1 &=& \frac{ A_1 \cos{2(l+g)} - A_2 \cos{2(l-g)}}{2 r}   = \frac{\sqrt{8 S}}{2 r} \,  \frac{\partial x_1}{\partial l}, \nonumber  \\
  X_2 &=& \frac{A_1 \sin{2 (l+g)} + A_2 \sin{2(l-g)}}{2 r}  = \frac{\sqrt{8 S}}{2 r} \,  \frac{\partial x_2}{\partial l},  \label{XLKS}\\
  X_3 &=& \frac{-B_1 \sin{2(l+\lambda)} + B_2 \sin{2(l-\lambda)}}{2 r}   = \frac{\sqrt{8 S}}{2 r} \,  \frac{\partial x_3}{\partial l}, \nonumber
\end{eqnarray}
where
\begin{equation}\label{rLKS}
    r = \frac{  L - B_1 \cos{2(l+\lambda)} - B_2 \cos{2 (l-\lambda)}}{  \sqrt{8 S}}.
\end{equation}
Finally, the `time deputy' variable $x^\ast = t$ is linked with the formal time-like variable $s$ through
\begin{equation}\label{stox}
    x^\ast = s - \frac{B_1 \sin{2(l+\lambda)} + B_2 \sin{2(l-\lambda)}}{4 S} = s - \frac{1}{\sqrt{8 S}} \frac{\partial r}{\partial l}.
\end{equation}
We have skipped the explicit expression of the KS variables, because it can be immediately obtained from the substitution
of (\ref{LKS:1}) into (\ref{LT1}-\ref{LT4}).

Two features of the above expressions for $\qq{x}$, $\qq{X}$, and $x^\ast$ deserve special attention. First, none of them
depends on $\gamma$, which means that any dynamical system primarily defined in terms of $\vv{x}$, $\vv{X}$, and time,
conserves the value of $\Gamma$. Secondly, the expressions for the Cartesian coordinates and momenta in the extended phase space
do not depend on the particular choice of $\alpha(S)$ and $\omega(S)$; the choice affects only the form of the Hamiltonian $\cM$.

\section{LKS variables and orbital elements}
\label{interp}

Let us interpret the variables forming the LKS set -- first the momenta, and then their conjugate angles -- by showing their relation to
the Keplerian elements or the Delaunay variables.

\subsection{LKS momenta}

Comparing equations (\ref{X0}) and (\ref{XLKS}) one immediately finds that
\begin{equation}\label{Gam:int}
      \mathcal{J}(\qq{v},\qq{V}) =   \Gamma ,
\end{equation}
when $\vv{c}=\vv{e}_3$, so observing that $\mathcal{J}(\qq{v},\qq{V})=0$ is the fundamental assumption of the KS transformation since the time of \citet{Kust:64},
there is no other choice than $\Gamma=0$. Recalling the absence of its conjugate angle $\gamma$ in the Hamiltonian, $\Gamma=0$ is
the integral of motion.

The meaning of $G$ becomes clear once we find the pull-back of the orbital angular momentum $\vv{G}_\mathrm{o}$ by $\zeta$, obtaining
\begin{eqnarray}
   \vv{G}_\mathrm{o} &=& \vv{x} \times \vv{X}  \nonumber \\
    & = &
  \frac{1}{2} \left(   C_1 \sin{2(g+\lambda)} - C_2  \sin{2(g-\lambda)} \right) \vv{e}_1 \nonumber \\
   & & +  \frac{1}{2} \left( - C_1 \cos{2(g+\lambda)} + C_2  \cos{2(g-\lambda)} \right) \vv{e}_2 \nonumber  \\
   & & +  \frac{G}{2} \vv{e}_3   + \frac{\Gamma \vv{x}}{2 r}. \label{G:Kep}
 \end{eqnarray}
Thus, setting $\Gamma=0$, we find the momentum $G$ to be twice the projection of the orbital angular momentum on the third axis (i.e. twice the Delaunay action $H_\mathrm{o}$).
Whenever the Hamiltonian admits the rotational symmetry around $\vv{e}_3$, the momentum $G$ will be the first integral of the system.

Proceeding to the momentum $L$, we have to distinguish the pure Kepler problem and the perturbed one. In the former case,
we can set $\cM_0=0$ in equation (\ref{M0}), finding
\begin{equation}\label{L:Kep}
    L = \frac{4\mu}{\alpha \omega} = \frac{2 \mu}{\sqrt{2 S}},
\end{equation}
at $\Gamma=0$.  Moreover, in the pure Kepler problem, the momentum $S$ can be expressed in terms of the major semi-axis $a$ as $S=\mu/(2a)$, which justifies
the direct link between the values of $L$ and of the Delaunay action $L_\mathrm{o}$
\begin{equation}\label{L:Kep:a}
    L = 2 \sqrt{\mu a} = 2 L_\mathrm{o}.
\end{equation}
The two restrictive clauses of the previous sentence (`values' and `pure Kepler') deserve comments. Equation (\ref{L:Kep:a})
does not imply differential relations, because, for example,
$\partial \vv{x}/\partial L_\mathrm{o} \neq  2\partial \vv{x}/\partial L$ \citep[c.f.][]{DepWil:91}.
Moreover, the values of $L$ and $2 L_\mathrm{o}$ generally differ in a perturbed problem due to the fact that
$L_\mathrm{o}$ is always defined by $\cH_0$ alone, whereas the definition of LKS momentum $L$ depends on the complete Hamiltonian $\cH_0+\mathcal{R}$
through the value of $S=X^\ast$ (the latter fixed by the restriction to the manifold $\cH=0$).

Similar intricacies are met for the momentum $\Lambda$, which turns out to be related with the Laplace (eccentricity) vector $\vv{e}$,
or rather the Laplace-Runge-Lenz vector $\vv{J} = L_\mathrm{o} \vv{e}$, having the dimension of angular momentum.
In the pure Kepler problem, substituting $\Gamma=0$, we find
\begin{eqnarray}
    \vv{J} &=& L_\mathrm{o} \left( \frac{\vv{X} \times \vv{G}}{\mu} - \frac{\vv{x}}{r}\right) \nonumber \\
 &=& \frac{1}{2} \left(  C_1 \sin{2(g+\lambda)} + C_2  \sin{2(g-\lambda)} \right) \vv{e}_1 \nonumber \\
   & & -  \frac{1}{2} \left( C_1 \cos{2(g+\lambda)} + C_2  \cos{2(g-\lambda)} \right) \vv{e}_2 \nonumber  \\
   & & +  \frac{\Lambda}{2} \vv{e}_3. \label{J:Kep}
\end{eqnarray}
Thus the momentum $\Lambda$ has been identified as twice the projection of the Laplace-Runge-Lenz vector on
the third axis, yet this equality, using the property $2 S L_\mathrm{o}^2 = \mu^2$,
holds only in the pure Kepler problem. In the perturbed case, one
should refer to the general definition of $\vv{e}$ in terms of the KS variables \citep{BL:17}.

Closing the discussion of the momenta, let us collect the bounds on their values:
\begin{equation}\label{bnd}
L > 0, \quad |\Lambda| + |G| \leqslant L, \quad  \Gamma=0.
\end{equation}
By the construction,
the value of $L=L_{12}+L_{03}$ must be nonnegative; but $L=0$ implies the permanent location at the origin ($\vv{x}=\vv{X}=\vv{0}$),
so we exclude it. The momenta $\Lambda$ and $G$ may be either positive or negative, but the above inequality guarantees that all
coefficients in equations (\ref{coefs}) are real.

\subsection{LKS angles}

\label{ang:int}
As already mentioned, the angle $\gamma$ is a cyclic variable, absent in the pullback of any Hamiltonian $\cH$ by $\zeta$.
Actually, $\gamma$ is the `KS angle' parameterizing the the fibre of KS variables $(\qq{v},\qq{V})$ mapped into the same point in the $(\vv{x},\vv{X})$
phase space. Thus, unless we are interested in some topological stability issues \citep{RUP:16}, the angle can be ignored.

The only fast angular variable is $l$. As expected, its values in the pure Kepler problem are equal to a half of the orbital eccentric anomaly $E$.
Indeed,
\begin{equation}\label{Etol}
    \frac{\rd l}{\rd t} =  \frac{\rd l}{\rd \tau} \frac{\rd \tau}{\rd t} = \frac{\partial \cM_0}{\partial L} \frac{\alpha(S)}{4r}
    = \frac{\omega(S) \alpha(S)}{4 r} = \frac{\sqrt{2 S}}{2 r} = \frac{\mu}{2 L_\mathrm{o} r} = \frac{1}{2}\,\frac{\rd E}{\rd t},
\end{equation}
and both angles are equal to 0 at the pericentre. Once again, this direct relation does not survive the addition of the perturbation.
Nevertheless, it also reveals the nature of equation (\ref{stox}) as a generalized Kepler's equation.

The two remaining angles are more unusual. A quick look at equations (\ref{G:Kep}) and (\ref{J:Kep})
might suggest that the role of $g$ and $\lambda$ in $\vv{G}_\mathrm{o}$ and $\vv{J}$ is similar. But if the norms of the vectors are evaluated,
one finds
\begin{eqnarray} 
  G_\mathrm{o} &=& \frac{1}{2} \sqrt{ G^2 + C_1^2+C_2^2 - 2 C_1 C_2 \cos{4 \lambda} } = L_\mathrm{o} \sqrt{1-e^2}, \label{Gnorm}  \\
  J &=& \frac{1}{2} \sqrt{  \Lambda^2 + C_1^2+C_2^2 + 2 C_1 C_2 \cos{4 \lambda} } = L_\mathrm{o} e. \label{Jnorm}
\end{eqnarray}
The absence of $g$ proves it to be some rotation angle; the presence of $\lambda$ means that this angle plays a different role
(and is somehow related with the eccentricity $e$).

More light is shed on this problem if we introduce the vectors
\begin{eqnarray} 
  \vv{M} &=& \frac{\vv{J}+\vv{G}_\mathrm{o}}{2} = \frac{C_1}{2} \sin{2(g+\lambda)} \, \vv{e}_1
  - \frac{C_1}{2} \cos{2(g+\lambda)} \, \vv{e}_2 + \frac{\Lambda+G}{4}\,\vv{e}_3, \label{Mv}  \\
  \vv{N} &=& \frac{\vv{J}-\vv{G}_\mathrm{o}}{2} =  \frac{C_2}{2} \sin{2(g-\lambda)} \, \vv{e}_1
  - \frac{C_2}{2} \cos{2(g-\lambda)} \, \vv{e}_2 + \frac{\Lambda-G}{4}\,\vv{e}_3. \label{Nv}
\end{eqnarray}
These are essentially the so-called Cartan or Pauli vectors \citep{Cordani}, except that we use the sign of
$\vv{N}$ opposite to the usual convention.
Both the vectors have the same norm $M=N= L/4 = L_\mathrm{o}/2$, and lie either in the plane perpendicular to orbit,
or along a degenerate radial orbit direction. The angle $\theta$ they form depends on the eccentricity alone, because
\begin{equation}\label{theta}
    \cos{\theta} = \frac{\vv{M} \cdot \vv{N}}{M N} = \frac{J^2-G^2_\mathrm{o}}{L_\mathrm{o}^2} =  2 e^2 - 1,
    \qquad \sin{\theta} = 2 e \sqrt{1-e^2}.
\end{equation}
Obviously, $\theta$ is the upper bound for the angle $\theta'$ between the projections of the Cartan vectors on
the coordinate plane $(x_1,x_2)$
\begin{equation}
\label{MNp}
\vv{M}' = \vv{M} - \frac{\Lambda+G}{4} \vv{e}_3, \qquad
\vv{N}' = \vv{N} - \frac{\Lambda-G}{4} \vv{e}_3.
\end{equation}
Using equations (\ref{Mv}), (\ref{Nv}), and (\ref{MNp}), one finds
\begin{equation} \label{ctp}
\cos{\theta'} = \frac{\vv{M}' \cdot \vv{N}'}{M' N'} =   \cos{4 \lambda}.
\end{equation}
Let us make $\theta'$ an oriented angle by postulating that it is measured from $\vv{N}'$ to $\vv{M}'$, counterclockwise.
Then its sine is given by
\begin{equation}\label{stp}
\sin{\theta'} = \frac{(\vv{N}' \times \vv{M}') \cdot \vv{e}_3 }{N' M'} =   \sin{4 \lambda}.
\end{equation}
Thus we have identified the angle $\lambda$ as the quarter of the angle between the projections of the Cartan
vectors on the reference plane $(x_1,x_2)$, measured from $\vv{N}'$ to $\vv{M}'$.
Finding $\theta'$ from the eccentricity-dependent $\theta$ involves orbital inclination and
the argument of pericentre, which means that $\lambda$ is a function of $e$, $I$, and $\omega_\mathrm{o}$.

Interestingly, whenever the argument of pericentre $\omega_\mathrm{o}$ exists, the statement $\sin{4\lambda}=0$
means $\cos{\omega_\mathrm{o}}=0$. Thus, any $\lambda = k\pi/4$ refers to $\omega_\mathrm{o}=\pi/2$ or $\omega_\mathrm{o}=3\pi/2$.

Once we have interpreted $\lambda$, the meaning of $g$ comes out of equations (\ref{Mv}) and (\ref{Nv}):
let us create the sum of normalized vectors $\vv{M}'/\|\vv{M}'\|+\vv{N}'/\|\vv{N}'\|$ and let us rotate the resulting vector
by $\pi/2$ counterclockwise, obtaining
\begin{equation}
  \vv{M}_\mathrm{m} =  2 \cos{2\lambda} \left( \cos{2g} \, \vv{e}_1 + \sin{2g} \, \vv{e}_2 \right).
\end{equation}
This formula suggests that $g$ is a half of the longitude of $\vv{M}_\mathrm{m}$, or of $-\vv{M}_\mathrm{m}$, depending on the sign
of $\cos{2\lambda}$. Whichever the case, changing the value of $g$ we perform a simultaneous rotation of both
$\vv{N}'$ and $\vv{M}'$ by the same angle. Indirectly, it means the rotation of the orbital plane
(if it exists) around the third axis, which makes $g$ a relative of the ascending node longitude.

\subsection{Special orbit types}

\begin{table} 
\caption{Particular orbits and their relation to the LKS variables.}
\label{tab:1}     
\begin{tabular}{lll}
\hline\noalign{\smallskip}
Orbit type & LKS variables & Undetermined angles  \\
\noalign{\smallskip}\hline\noalign{\smallskip}
Generic circular & $\Lambda=0, \, 0<|G|<L, \, \lambda =(2k+1)\frac{\pi}{4}$  & none \\
Circular, polar & $\Lambda=0, \, G=0, \,\lambda =(2k+1)\frac{\pi}{4}$  & none \\
Circular, equatorial & $\Lambda=0, \, |G|=L$ & $l,g,\lambda$ \\
Generic radial & $G=0,\, 0 < |\Lambda|<L,\, \lambda =k \frac{\pi}{2}$ & none \\
Radial, equatorial & $G=0,\, \Lambda=0,\, \lambda =k \frac{\pi}{2}$ & none \\
Radial, polar & $G=0,\, |\Lambda|=L $ & $l,g,\lambda$ \\
Generic equatorial & $\Lambda=0,\, 0 < |G|<L,\, \lambda =k \frac{\pi}{2}$ & none \\
\noalign{\smallskip}\hline
\end{tabular}
\end{table}

Let us inspect how some specific orbit types are mapped onto the LKS variables. The discussion is
restricted to the elliptic orbits ($0 \leqslant e \leqslant 1$) in the pure Kepler problem.

\subsubsection{Circular orbits}

Circular orbits, having $e=0$, are characterized by $\vv{J}=\vv{0}$, hence all must possess $\Lambda=0$, since $2 \Lambda = \vv{J}\cdot \vv{e}_3$.
Then, the norm of the Laplace-Runge-Lenz vector (\ref{Jnorm}) simplifies, thanks to $C_1=C_2=\sqrt{L^2-G^2}/2$, and equating its square to 0 we find the condition
\begin{equation}\label{con:1}
   4 J^2 =  \left(L^2-G^2\right) \, \left(\cos{2\lambda}\right)^2 = 0.
\end{equation}
Setting $\lambda = (2k+1) \pi/4$, $k \in \mathbb{Z}$, leads to generic circular orbits with the inclination $I = \arccos{(G/L)}$, including
circular polar orbits when $G=0$. However, if $|G|=L$, then the first factor is null regardless of $\lambda$. This is the case of
circular orbits in the `equatorial plane' $(x_1,x_2)$: prograde for $G=L$, or retrograde for $G=-L$.

The values of $\lambda$ mentioned above well coincide with the interpretation from section~\ref{ang:int}. In circular orbits,
the Cartan vectors $\vv{N}$ and $\vv{M}$ are collinear and opposite, thus the angle $\theta=\pi$, and its projection $\theta'$
remains $\pm \pi$ as long as the orbit is not equatorial. Thus $\lambda = \theta'/4 = \pm \pi/4$, plus any multiple of  $(2\pi)/4$.

Another explanation of the LKS variables for $e=0$ can be given by inspecting the Lissajous ellipses in Fig.~\ref{fig:2}.
The orbital distance $r$ is the sum of $\rho^2_{03} =v_0^2+v_3^2$ and
$\rho^2_{12}  = v_1^2+v_2^2$, both divided by $\alpha$. In order to secure a constant $r= (\rho^2_{12}+\rho^2_{03})/\alpha$, it is not necessary that
both $\rho_{ij}$ are constant; enough if they oscillate with the same amplitude and a phase shift of $\pm \pi/2$.
Equal amplitudes result from $L_{12}=L_{03}$ (because $G_{12}=G_{03}$ by $\Gamma=0$), hence $\Lambda= L_{12}-L_{03}=0$. The phase shift condition is given by
$l_{12}-l_{03} = 2 \lambda  = (2k+1) \pi/2$, which means the values of $\lambda$ as above.

The case of constant $\rho_{ij}$, mentioned above, should be related with some special kind of a circular orbit. Indeed, since it needs $L_{12}=|G|/2=L_{03}$, i.e.
two circles of equal radii in Fig.~\ref{fig:2}, we obtain the circular equatorial orbits with $\Lambda=0$ and $|G|=L$ (prograde or retrograde,
depending on the sign of  $G$). Observe that due to the lack of distinct semi-axes in the two circles, the angles $l_{ij}$, and $g_{ij}$ are undefined,
and so are $l$, $g$, $\gamma$, and $\lambda$. But still one can use properly defined `longitudes'
$l+g$ or $l-g$ in the prograde, and retrograde cases, respectively -- at least until some `virtual singularities' appear \citep{Hen:74}.
In terms of the Cartan vectors, $\vv{M}= -\vv{N} = \vv{G}_\mathrm{o}$, so $\vv{M}'=\vv{N}'=\vv{0}$, making the angles $g$ and $\lambda$ undetermined.

\subsubsection{Radial orbits}

Rectilinear (radial) orbits require $\vv{G}_\mathrm{o}=\vv{0}$, hence $G=0$ and $\vv{G}_\mathrm{o} \cdot \vv{G}_\mathrm{o} = 0$.
According to equations~(\ref{coefs}), $G=0$ means $C_1=C_2= \sqrt{L^2-\Lambda^2}/2$, wherefrom equation (\ref{Gnorm}) implies
\begin{equation}\label{con:2}
   4 G_\mathrm{o}^2 =  \left(L^2-\Lambda^2\right) \, \left(\sin{2\lambda}\right)^2 = 0.
\end{equation}
Regardless of $\lambda$, it is satisfied by $|\Lambda|=L$, i.e. by polar radial orbits with $\vv{J} = (\Lambda/2)\,\vv{e}_3$.
For all other directions of the Laplace-Runge-Lenz vector, radial orbits need $\lambda = k \pi/2$, where $k \in \mathbb{Z}$;
this time $\Lambda$ can be arbitrary, with $\Lambda=0$ indicating an equatorial radial orbit. In terms of the Cartan vectors,
$\vv{G}_\mathrm{o}=0$ means $\vv{N}=\vv{M}$, so their angle $\theta=0$ is projected as $\theta'=0+2 k\pi$, which (divided by 4)
gives the above values of $\lambda$.

In terms of the Lissajous ellipses in $(v_1,v_2)$ and $(v_0, v_3)$ planes from Fig.~\ref{fig:2}, $G=0$ means that both degenerate into straight segments.
The motion along the segments must obey $l_{12}=l_{03}+k\pi$, to guarantee that $v_0=v_1=v_2=v_3=0$ at the same epoch. The direction of $\vv{x}(\qq{v})$
is determined by the difference of lengths of the two segments: equatorial orbits result if the segments have the same length, whereas polar orbits
require that one of the segments collapses into a point. In the latter case, $l$ and $\lambda$ are undetermined, but $l+\lambda=l_{02}$ or $l-\lambda=l_{03}$
retain a well defined meaning for an appropriate sign of $\Lambda$. Problems with the definition of $g_{ij}$ due to the vanishing minor axes
are only apparent, because they can be solved by an alternative definition: instead of `position angle of the minor semi-axis', one can equally well say
`position angle of the major semi-axis minus $\pi/2$'.

\subsubsection{Equatorial orbits}

Since the Laplace-Runge-Lenz vector lies in the plane $(x_1,x_2)$ for the equatorial orbits, all they must have $\Lambda=0$, whence
$C_1=C_2=\sqrt{L^2-\Lambda^2}/2$. Moreover, $G_\mathrm{o}^2=(G/2)^2$, which leads to the condition
\begin{equation}\label{con:3}
   4 (G_\mathrm{o}^2-G^2) =  \left(L^2-G^2\right) \, \left(\sin{2\lambda}\right)^2 = 0.
\end{equation}
The case of $|G|=L$ brings us back to the circular equatorial orbits, already discussed. Other values of $G$ require $\lambda = k \pi/2$, where $k \in \mathbb{Z}$.
These are the same values as in the case of radial orbits, which makes sense, because $G=0$ should bring us to the radial equatorial orbit.

For an elliptic ($e\neq 0$) equatorial orbit, the Cartan vectors $\vv{N}$ and $\vv{M}$ may form different angles $\theta$, but since they lie in a polar plane,
the projection of these angle is always $\theta'=0$, exactly as in the radial orbit case -- thus the same values of $\lambda$.

The two Lissajous ellipses in Fig.~\ref{fig:2} must have the same semi-axes, and $l_{12}=l_{03}+k\pi$. This is necessary to obtain $v_1^2+v_2^2=v_0^2+v_3^2$,
which guarantees $x_3=0$ for all epochs, according to equation (\ref{KSgen}) in the KS3 setup.

\subsubsection{Polar orbits}

Polar orbits are generically indicated by the simple condition $G=0$. It is only in the special cases where the angle $\lambda$ comes into play:
circular polar orbits ($\Lambda=0$) need $\lambda=(2k+1)\pi/4$, whereas radial polar orbits $(|\Lambda|=L)$ are the ones where $\lambda$ is undetermined.
Since $G=0$, both Lissajous ellipses degenerate into segments, but their lengths may be different, and the phase shift arbitrary.

\subsubsection{Singularities}
\label{sing}
Inspecting specific types of orbits we met the situations, where $\lambda$ and $g$ become undetermined: circular equatorial orbits
with $|G|=L, \Lambda=0$ and rectilinear polar orbits with $|\Lambda|=L, G=0$. These four points are the vertices of the square
on the $(G,\Lambda)$ plane defined by the constraint $|G|+|\Lambda| \leqslant L$. However, all four edges of the square
leave the angles undetermined. This is related to the fact, that:
\begin{itemize}
\item[a)] $L=G+\Lambda$ (upper right edge in Fig.~\ref{bifur}) means $L_{03}=G_{03}$, i.e. prograde circular motion on $(v_0,v_3)$ plane with undetermined $l_{03}$ and $g_{03}$ (but $l_{03}+g_{03}$ is well defined),
\item[b)] $L=-G+\Lambda$ (upper left edge in Fig.~\ref{bifur}) means $L_{03}=-G_{03}$, i.e. retrograde circular motion on $(v_0,v_3)$ plane with undetermined $l_{03}$ and $g_{03}$ (but $l_{03}-g_{03}$ is well defined),
\item[c)] $L=G-\Lambda$ (lower right edge in Fig.~\ref{bifur}) means $L_{12}=G_{12}$, i.e. prograde circular motion on $(v_1,v_2)$ plane with undetermined $l_{12}$ and $g_{12}$ (but $l_{12}+g_{12}$ is well defined),
\item[d)] $L=-G-\Lambda$ (lower left edge in Fig.~\ref{bifur}) means $L_{12}=G_{12}$, i.e. retrograde circular motion on $(v_1,v_2)$ plane with undetermined $l_{12}$ and $g_{12}$ (but $l_{12}-g_{12}$ is well defined).
\end{itemize}
The Keplerian orbits obtained by mapping the edges of the $(G,\Lambda)$ square onto
$\vv{J}$ and $\vv{G}_0$ or $\vv{x}$ and $\vv{X}$, have $e=\sin{I}$, and $\sin\omega_\mathrm{o}=\pm 1$. Thus the vertices in Fig.~\ref{bifur}
are $\sin{I}=e=0$ (left and right) and $\sin{I}=e=1$ (top and bottom).
Along the edges, half of the coefficients (\ref{coefs}) does vanish, and
one of the vanishing coefficients is always $C_1$ or $C_2$, which implies that either $\vv{M}'$ or $\vv{N}'$
is a null vector, so the angles $\lambda$ and $g$ become undefined.

\section{Application to the Lidov-Kozai problem}

\label{Koza}
\subsection{Derivation of the secular model}

In order to test the LKS variables in a nontrivial astronomical problem, let us revisit the Lidov-Kozai resonance
arising in the artificial satellites theory \citep{Lidov:62} or asteroid dynamics \citep{Kozai:62}.
In this already classical problem, the Keplerian motion of a small body (a satellite or an asteroid)  around a central mass with the gravitational parameter $\mu$ (a planet or the Sun)
is influenced by a distant perturber with the gravitational parameter $\mu'$ (the Sun or a planet, respectively). The origin of the reference frame is attached to
the central mass, the plane $(x_1,x_2)$ coincides with the orbital plane of the perturber, and the third axis basis vector $\vv{e}_3$ is directed along the
angular momentum of the perturber. Further, let us assume that the perturber moves on a circular orbit with the mean motion $n_\mathrm{p}$, so its position vector is
\begin{equation}\label{rp}
    \vv{r}_\mathrm{p} = a_\mathrm{p} \cos n_\mathrm{p}t \,\vv{e}_1 + a_\mathrm{p} \sin n_\mathrm{p}t\, \vv{e}_2.
\end{equation}
Compared to the small body, whose position vector is  $\vv{x}$, the perturber is distant, i.e. $||\vv{x}||/||\vv{r}_\mathrm{p}|| = r/a_\mathrm{p}$
is small enough to approximate the perturbing function by the second degree Legendre polynomial term. Thus we obtain the problem
with the Hamiltonian $\cH$ from equation (\ref{Ham}) with the perturbation
\begin{equation}\label{RLK}
    \mathcal{R} = - \frac{\mu_\mathrm{p} r^2}{a_\mathrm{p}^3}\,P_2(\vv{x}\cdot \vv{r}_\mathrm{p}/(r a_\mathrm{p})).
\end{equation}
The perturbation is time-dependent, so -- after substituting (\ref{rp}) -- we replace $t$ by its formal twin $x^\ast$, obtaining
\begin{equation}\label{R1}
    \mathcal{R} = - \frac{\mu_\mathrm{p}}{4 a_\mathrm{p}^3} \left[
    r^2-x_3^2+3 (x_1^2-x_2^2) \cos{2 n_\mathrm{p} x^\ast} + 6 x_1 x_2 \sin{2 n_\mathrm{p} x^\ast}
    \right].
\end{equation}
Choosing $\omega=1$, and $\alpha = \sqrt{8 X^\ast} = \sqrt{8 S}$, we apply the LKS transformation, setting $\Gamma=0$,
because we are not interested in the evolution of the KS angle $\gamma$.
The resulting Hamiltonian (\ref{M}) is
\begin{equation}\label{MLK}
    \cM = \cM_0 + \mathcal{Q} = 0,
\end{equation}
with the Keplerian part
\begin{equation}\label{M0LK}
    \cM_0 = L - \frac{2 \mu}{\sqrt{2 S}}.
\end{equation}
For a while, the perturbation  $\mathcal{Q}$ will be given in an intermediate form without the explicit substitution
of the LKS variables into $\vv{x}$ and $r$, which leads to a relatively concise form
\begin{equation}\label{QLK:1}
    \mathcal{Q} = - \frac{\mu_\mathrm{p} r}{32 a_\mathrm{p}^3 \sqrt{2 S}} \left[
    r^2-x_3^2+3 (x_1^2-x_2^2) \cos{\left(2 n_\mathrm{p} s - \sigma\right)} + 6 x_1 x_2 \sin{\left(2 n_\mathrm{p} s - \sigma\right)}
    \right],
\end{equation}
where, according to equation (\ref{stox}),
\begin{equation}\label{sig}
  \sigma = \frac{n_\mathrm{p}}{2 S} \left(   B_1 \sin{2(l+\lambda)} + B_2 \sin{2(l-\lambda)} \right).
\end{equation}
By the choice of $\alpha$, the Sundman time $\tau$ is dimensionless and the unperturbed motion gives
\begin{eqnarray} 
  \frac{\rd l}{\rd \tau}  &=& \frac{\partial \cM_0}{\partial L} = 1,  \label{dldt}\\
  \frac{\rd s}{\rd \tau} &=& \frac{\partial \cM_0}{\partial S} = \frac{\mu}{\sqrt{2 S^3}},
\end{eqnarray}
with all the remaining variables constant.
Solving (\ref{dldt}) we find  $l=\tau+l_0$. The value of $S$ is set to give $\cM=0$, but
ignoring the contribution of $\mathcal{Q}$ we may estimate that $s \approx \tau/n$, where $n$ is the Keplerian mean motion.

According to the standard Lie transform method \citep[e.g.][]{FM:07}, the mean variables can be introduced by a nearly canonical
transformation that converts $\cM$ into $\mathcal{N}=\mathcal{N}_0+\mathcal{Q}'$, with $\mathcal{N}_0 = \mathcal{M}_0$ and
$\mathcal{Q}'$ being constant along the phase trajectory generated by $\mathcal{N}_0$. Up to the first order, the
new perturbation $\mathcal{Q}'$ is simply the average of $\mathcal{Q}$ with respect to $\tau$, assuming $l=\tau+l_0$ and
$s=\tau/n$.

Since the perturber has been assumed distant, its mean motion $n_p$ is small compared to $n$ and both frequencies can be treated as irrational; even if they are not, the resonance will occur in high 
degree harmonics with practically negligible amplitudes.
In these circumstances, any product of sine or cosine  of $2n_\mathrm{p} s = 2 (n_\mathrm{p}/n) \tau$ with a function which is either constant
or $2\pi$-periodic in $\tau$, has the zero average.\footnote{The general definition of the
average for a function $f(\tau)$ is  $\lim_{\tau\rightarrow \infty} \tau^{-1} \int_0^\tau f(\tau')\rd \tau'$,
so its value for a quasi-periodic function is null. When $f(\tau)$ is $T$-periodic, this definition simplifies to the standard $T^{-1} \int_0^T f(\tau)\rd \tau$.} Then $\mathcal{Q}'$ simplifies to
\begin{equation}\label{QLK:2}
    \mathcal{Q}' = - \frac{\mu_\mathrm{p} }{64 \pi a_\mathrm{p}^3 \sqrt{2 S}}\int_0^{2\pi} \left[
    r^3- r x_3^2
    \right] \rd l.
\end{equation}
Thus we obtain the first order approximation of the secular system
\begin{eqnarray}\label{Nf}
    \mathcal{N} & = & L - \frac{2 \mu}{\sqrt{2 S}}
  - \frac{\mu_\mathrm{p}\, L}{1024\, a_\mathrm{p}^3\, S^2}
    \left( L^2 -6 \Lambda^2 + 6 C_1 C_2 \cos{4\lambda} \right) = 0, \\
    C_1 C_2 & = & \frac{1}{4}  \sqrt{\left(L^2-(G-\Lambda)^2\right)\,\left(L^2-(G+\Lambda)^2\right)},
\end{eqnarray}
where the mean variables should be given different symbols, but we adhere to a widespread habit of distinguishing the  mean and the osculating
variables by context. The following study of motion generated by $\mathcal{N}$ will refer only to the mean variables, so no confusion should occur.

Both $\mathcal{N}$ and the classical secular Hamiltonian of the Lidov-Kozai problem share the same property: they are
reduced to 1 degree of freedom. In our case it is the canonically conjugate pair $(\lambda,\Lambda)$ instead of the usual Delaunay pair of
the argument of pericentre and the angular momentum norm. All other momenta are constant and will be treated as parameters.
However, there is a fundamental difference between our formulation and the classical approach: the equations of motion for $\lambda$ and $\Lambda$
are not singular for most of the radial orbits.

\subsection{Secular motion and equilibria}

\begin{figure}
	\includegraphics[width=\textwidth]{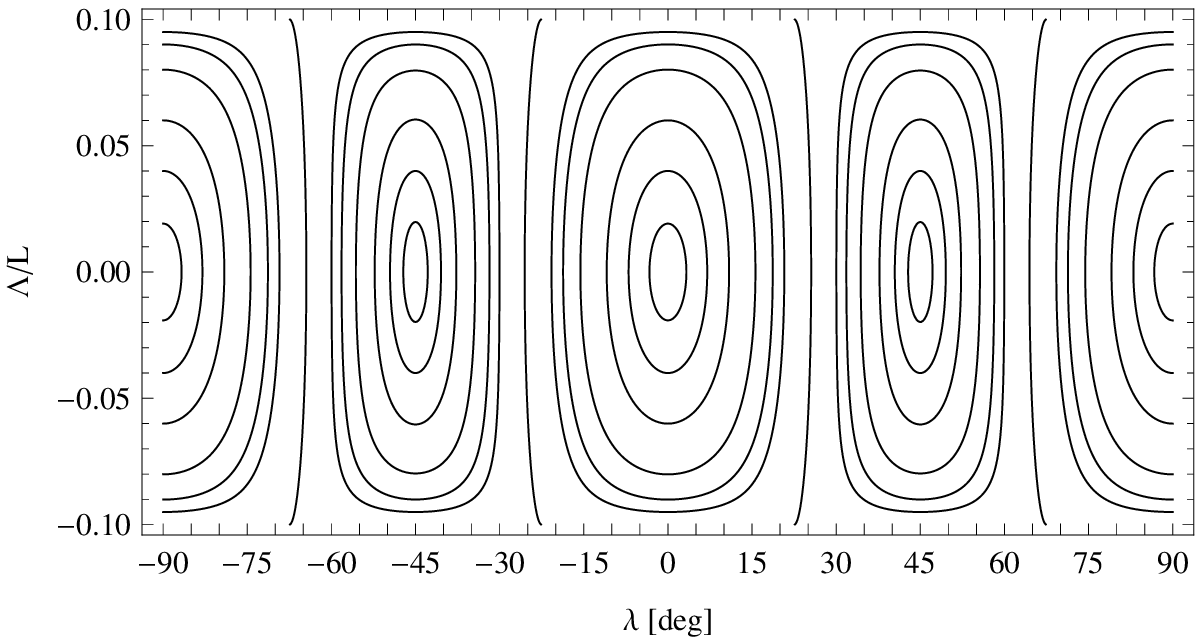}\\
    \includegraphics[width=\textwidth]{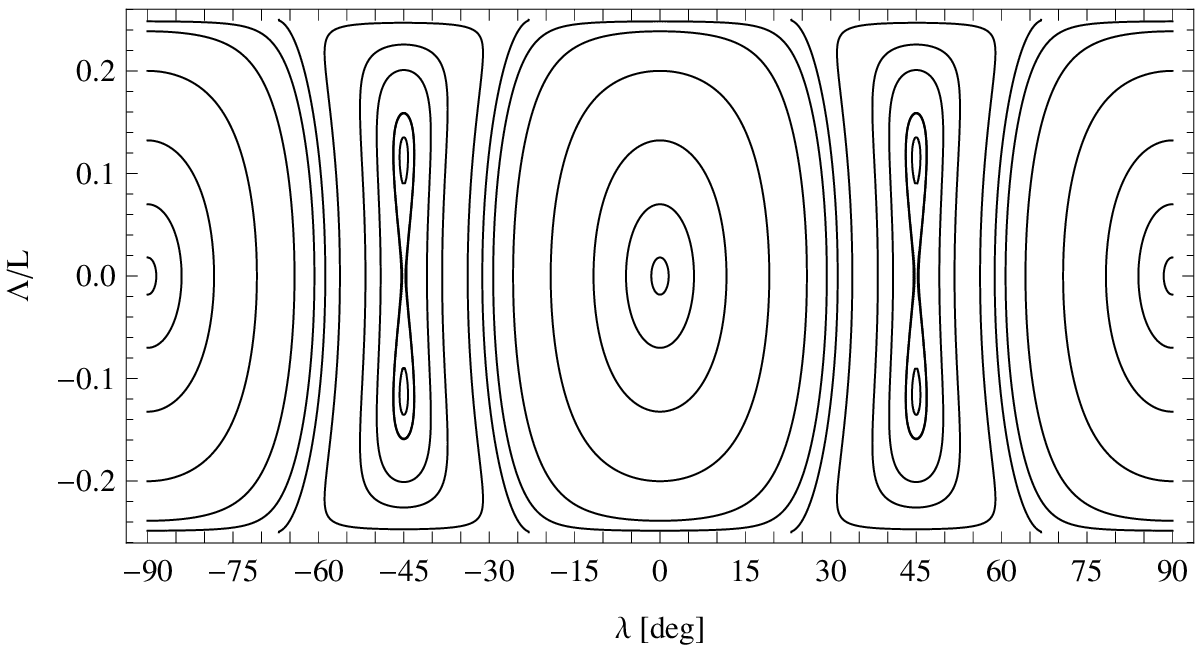}\\
    \includegraphics[width=\textwidth]{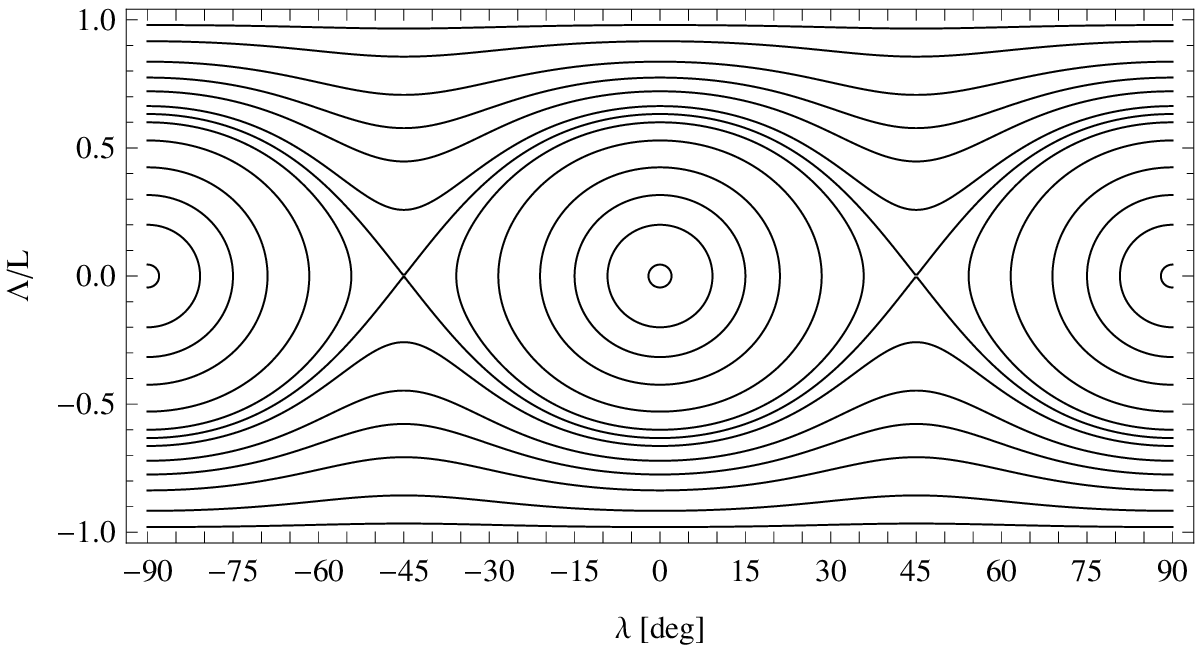}
    \caption{Integral curves of the regularized Lidov-Kozai problem on the $(\lambda,\Lambda)$ phase plane. Top: $G = 0.9\,L$, middle: $G=0.75\,L$,
    bottom: $G=0$.}
    \label{LKpp}
\end{figure}

Let us set
\begin{equation}\label{Bdef}
    B = \frac{3 \mu_\mathrm{p} L}{1024 a_\mathrm{p} S^2}.
\end{equation}
The equations of motion derived from (\ref{Nf}) are
\begin{eqnarray} 
  \frac{\rd \lambda}{\rd \tau} &=& \frac{\partial \mathcal{N}}{\partial \Lambda} =
  B \,\Lambda \left( 4 + \frac{L^2+G^2-\Lambda^2}{4 C_1 C_2} \cos{4 \lambda} \right), \label{eq:lam} \\
   \frac{\rd \Lambda}{\rd \tau} &=& -\frac{\partial \mathcal{N}}{\partial \lambda} = - 8 B C_1 C_2
   \sin{4 \lambda}.
\end{eqnarray}
Integral curves of this system are plotted in Fig.~\ref{LKpp} for three values of $G$: $0.9L$, $0.75 L$ and $0$. The phase plane has been clipped to $-\pi \leqslant \lambda \leqslant  \pi$,
because the reaming range of $\lambda$ is a simple duplication of the plotted phase portrait.

Referring to Table~\ref{tab:1}, one can check that a generic radial orbit does not introduce a singularity
into equations (\ref{eq:lam}). Indeed, $G=0$, and $\lambda= k \pi/2$
result in a well defined
\begin{equation}\label{rec:1}
 \frac{\rd \lambda}{\rd \tau} = 5 B \Lambda , \quad \frac{\rd \Lambda}{\rd \tau} = 0.
\end{equation}
It means that a radial orbit is not en equilibrium, unless $\Lambda=0$, which is exactly the case of a radial orbit in the equatorial plane.
Observing that for $G=0$ the points $(\lambda= k\pi/2,\Lambda=0)$ are well defined local minima of $\mathcal{N}$,
we are able to state that radial orbits in the orbital plane of the perturber are stable\footnote{The word `stable' is a bit paradoxical in
this context, because it means that the motion starting in such an orbit will inevitably end up in collision
with the central body.} equilibria.The bottom panel of  Figure~\ref{LKpp} confirms this observation: the points $(0,0)$, $(90^\circ,0)$, and $(-90^\circ,0)$ are surrounded by
closed, oval shape contours. Intersection of any of the integral curves plotted in the bottom panel with the vertical lines $\lambda = 0$, or $\lambda=\pm 90^\circ$
marks a temporary passage through the radial orbit degeneracy.

As far as the polar radial orbits (with $|\Lambda|=L$) are concerned,
equations (\ref{eq:lam}) become singular, but this singularity is purely geometrical. Such orbits should be located at the upper and lower edges of
 the bottom panel in Fig.~\ref{LKpp}, where $\lambda$ is undetermined. But since the integral curves approaching the edges become parallel to them, one should expect
 that polar radial orbits are stable equilibria (which is actually the case, if the analysis is performed in terms of vectors $\vv{G}_\mathrm{o}$ and $\vv{J}$,
 or simply observing that for $G=0$ the Hamiltonian $\mathcal{N}$ has the local maxima at $\Lambda = \pm L$ regardless of the value of $\lambda$).

Actually, The presence of $\Lambda$ as a factor of the first of equations (\ref{eq:lam})
means that for any value of $|G| \neq L$, the equilibria exist at
$(\lambda= j\,\pi/4,\Lambda=0)$, as seen in Fig.~\ref{LKpp}. For even $j=2k$, the equilibria refer to equatorial orbits with the eccentricity depending on $G$ through
$e=\sqrt{1-(G/L)^2}$. It is easy to check the they are the local minima of the Hamiltonian $\mathcal{N}$, hence the equatorial orbits are stable.
The circular equatorial case with $|G|=L$  is problematic, because then the upper and lower limits of $\Lambda$ merge, and in order to prove that
these are actually the stable equilibria one has to resort to the analysis of $\vv{G}_\mathrm{o}$ and $\vv{J}$ vectors.

For odd $j=(2k+1)$, the equilibria are circular orbits with inclinations depending on $G$ (equatorial if $G=0$,
prograde for $G>0$ and retrograde when $G<0$). Their stability depends on the ratio $G/L$. Unlike in the Delaunay chart, variational equations can be
formulated directly in the phase plane of $(\lambda,\Lambda)$, leading to the eigenvalues that are pure imaginary for $(G/L)^2 > 3/5$. Thus
circular orbits are stable for inclinations below $I_1=\arccos{\sqrt{3/5}} \approx 39^\circ\!.23$ and above  $I_2=\arccos{-\sqrt{3/5}} \approx 140^\circ\!.77$.
At these critical values a bifurcation occurs: when  $(G/L)^2 < 3/5$ circular orbits become unstable and two stable equlibria are created
at $(\lambda= (2k+1)\,\pi/4, \Lambda = \pm \Lambda_\mathrm{c})$ (see the middle panel of Fig.~\ref{LKpp}). Recall that, in general case of inclined, elliptic orbits, this value of $\lambda$
means the argument of pericentre equal $\pi/2$ or $3\pi/2$. The value of $\Lambda_\mathrm{c}$ is the root of the first of equations (\ref{eq:lam}) with
$\Lambda \neq 0$ and $\cos{4\lambda}= -1$, i.e.
\begin{equation}\label{cLc}
    4 - \frac{L^2+G^2-\Lambda^2}{4 C_1 C_2} = 0,
\end{equation}
leading to
\begin{equation}\label{Lc}
    \Lambda_\mathrm{c} = L \sqrt{1-\frac{8 |G|}{\sqrt{15} L} + \left(\frac{G}{L}\right)^2}.
\end{equation}
These are the classical equilibria of the Lidov-Kozai problem -- the only ones that can be analyzed directly in the Delaunay variables. In terms
of the orbital elements, equation (\ref{Lc}) is equivalent to the well known condition \citep{Lidov:62}
\begin{equation}\label{LcK}
    1-e^2 = \frac{3 (\cos{I})^2}{5}.
\end{equation}
The equilibria can be located in the middle panel of Fig.~\ref{LKpp} at $\lambda = \pm 45^\circ$, $\Lambda \approx \pm 0.115\,L$.

\begin{figure}
	\includegraphics[width=\textwidth]{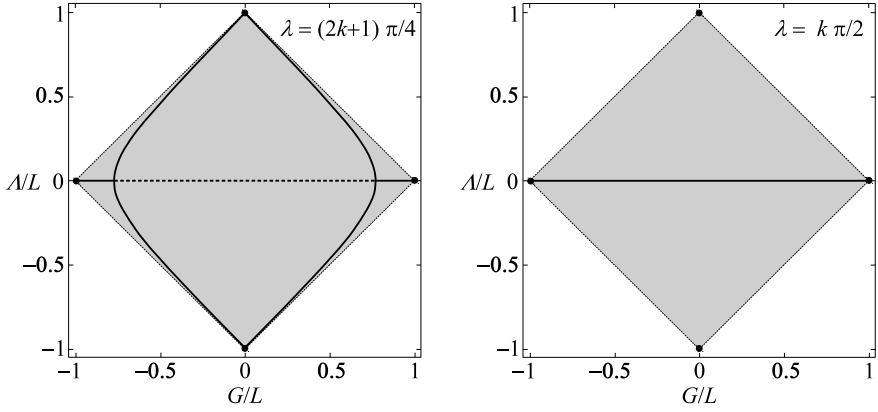}
    \caption{Equilibria in the regularized Lidov-Kozai problem. Grey square -- admissible region of $G$ and $\Lambda$.
    The case of $\lambda = (2k+1) \pi/4$: horizontal line -- circular orbits (solid line for stable, dashed for unstable equilibria),
    vertical curves -- stable classical  equilibria (\ref{LcK}). The case of $\lambda = k \pi/2$: horizontal line -- stable equatorial orbits.
    Solid circles at the vertices -- stable equilibria with undetermined $\lambda$: polar radial ($G=0$) and equatorial circular ($\Lambda=0$) orbits.}
    \label{bifur}
\end{figure}

Figure~\ref{bifur} shows all the equilibria and their stability, with the dashed lines marking the unstable equilibrium. The edges
of the $(G,\Lambda)$ square (upper and lower boundaries of the plots in Fig.~\ref{LKpp}) may not be attached to any of the values of $\lambda$,
but we added the black dots at the corners to show the stable equilibria of the special type as the natural limits of the
stable branches (solid lines).

It is not unusual that all action-angle-like variables with bounded momentum suffer from indeterminate angle at the boundary of its conjugate.
The LKS variables cannot be different, even if many cases, problematic in the Delaunay chart, have been located inside the boundaries of $\Lambda$.
For each value of $G \neq 0$ (and $|G| \neq L$) there exist integral curves passing through
both the extremes: $\Lambda = L - |G|$ and $\Lambda = -L+ |G|$. In the top or the middle panel of Fig.~\ref{LKpp} they are seen as four disjoint fragments;
for example, the two open curves approaching the edges at $\lambda \pm 22^\circ\!.5$ are the fragments of such an integral curve. There is no singularity in
these orbits (see Sect.~\ref{sing}) other than the indeterminacy of longitude at the poles of a sphere \citep{Dep:94}.

\section{Conclusions}
\label{conc}

While commenting a transformation due to Fukushima, \citet{Dep:94} observed that it amounts to swapping singularities, and immediately added
`This remark is not meant to diminish its practical merit, quite the contrary.' The LKS variables we have presented also `trade in singularities',
but the rule of trade we propose is to spare the radial, rectilinear orbits (except the polar ones) at the expense of some other types.
The exceptions include mostly a family of expendable, rank-and-file orbits with $e=\sin{I}$ and the lines of apsides perpendicular to the lines of nodes -- the cases
easily tractable without the KS regularization and unlikely to focus attention by becoming equilibria in typical problems of celestial mechanics.
More we regret the problems caused by the polar radial, and equatorial circular orbits. Nevertheless, we believe that more has been gained than lost.
Enough to enumerate the orbits that remain regular points in our chart: circular inclined,  equatorial elliptic, and all radial (except the polar ones).
Thanks to refraining from the use of orbital plane in their construction, the LKS variables are better fitted to study
highly elliptic orbits than any other action-angle set known to the authors.

The analysis of the quadrupole Lidov-Kozai problem in Section~\ref{Koza} suggests that the LKS variables may be a handy tool in the analysis of the
more problematic cases, like the eccentric, octupolar Lidov-Kozai problem. In the latter, the `orbital flip' phenomenon occurs: changing the direction of motion
with the passage through an equatorial rectilinear orbit phase \citep{LiNa:2011}. Previous attempts to discuss this phenomenon in terms of the
action-angle variables \citep[e.g][]{Sidor:18} faced the problems which may possibly be resolved with the newly presented parametrization.

Some of the readers might be sceptical about the unnecessary duplication of the phase space resulting from the LKS transformation $\zeta$.
Indeed, Fig.~\ref{LKpp} covers the whole phase space of in terms of the argument of pericentre $\omega_\mathrm{o}$, although it has been clipped to
the half range of $\lambda$. This feature can be trivially removed by means of a symplectic transformation $(\lambda,\Lambda) \rightarrow (2 \lambda, \Lambda/2)$,
and similarly for other conjugate pairs. We have not made this move in the present work for the sake of retaining the fundamental, angle-halving property of
both the Levi-Civita and the Kustaanheimo-Stiefel transformations. Avoiding factor 2 in the arguments of sines and cosines  in equations (\ref{xLKS}) and
(\ref{XLKS}), we would introduce the factor $\frac{1}{2}$ in the expressions for $\qq{v}$ and $\qq{V}$. Let us mention that the restriction of
the LKS transformation to $(\qq{v},\qq{V}) \rightarrow (l,g,h,\gamma,L,G,H,\Gamma)$ can be useful also in the studies of
perturbed, four degrees of freedom oscillators, not necessarily resulting from the KS transformation \citep[e.g.][]{CMF:15,vdM:16}.
In that case, unwanted spurious singularities may arise in course of the Birkhoff normalization, when the multiple of angle does not properly match
the power of action.

Having based the LKS variables upon the KS3 variant of the KS transformation, we do not exclude a possibility of performing a similar construction within the KS1 framework.
But then the $G$ and $\Lambda$ variables will be the projections of the angular momentum and the Laplace-Runge-Lenz vectors on the $x_1$ axis.
With such a choice, the Lidov-Kozai Hamiltonian (\ref{Nf}) would depend on both $g$ and $\lambda$, with the rotation symmetry hidden deeply
in some complicated function of all variables, instead of the obvious $G=\mbox{const}$.

%\begin{acknowledgements}
%If you'd like to thank anyone, place your comments here
%and remove the percent signs.
%\end{acknowledgements}
~\\
\noindent
\textbf{Compliance with ethical standards}\\
~\\
\textbf{Conflict of interest}{ The authors S. Breiter and K. Langner declare that they have no conflict of interest.}

% BibTeX users please use one of
\bibliographystyle{spbasic}      % basic style, author-year citations
\bibliography{LKS}   % name your BibTeX data base

\end{document}